\documentclass[sigconf]{acmart}

\usepackage{cite}
\usepackage{amsmath,amsfonts}
\usepackage{algorithmic}
\usepackage{graphicx}
\usepackage{textcomp}
\usepackage{xcolor}
\usepackage{multirow}       
\usepackage[utf8]{inputenc} 
\usepackage[T1]{fontenc}    
\usepackage{hyperref}       
\usepackage{url}            
\usepackage{booktabs}       
\usepackage{amsfonts}       
\usepackage{nicefrac}       
\usepackage{microtype}      
\usepackage{graphicx}
\usepackage{amsmath}
\usepackage{amsthm}
\newtheorem{theorem}{Theorem}

\newtheorem{lemma}{Lemma}
\newtheorem{definition}{Definition}

\usepackage{bbm}
\usepackage{bm}
\usepackage{mathrsfs}
\usepackage{enumitem}
\usepackage{mathtools} 
\usepackage{tikz} 
\usepackage{fancyhdr}
\usepackage[ruled,linesnumbered]{algorithm2e}

\AtBeginDocument{%
  \providecommand\BibTeX{{%
    \normalfont B\kern-0.5em{\scshape i\kern-0.25em b}\kern-0.8em\TeX}}}

\copyrightyear{2024}
\acmYear{2024}
\setcopyright{acmlicensed}\acmConference[CIKM '24]{Proceedings of the 33rd ACM International Conference on Information and Knowledge Management}{October 21--25, 2024}{Boise, ID, USA}
\acmBooktitle{Proceedings of the 33rd ACM International Conference on Information and Knowledge Management (CIKM '24), October 21--25, 2024, Boise, ID, USA}
\acmDOI{10.1145/3627673.3679536}
\acmISBN{979-8-4007-0436-9/24/10}

\begin{document}

\title{Interpretable Triplet Importance for Personalized Ranking}


\author{Bowei He}
\affiliation{%
  \institution{City University of Hong Kong}
  \country{Hong Kong SAR}
}
\email{boweihe2-c@my.cityu.edu.hk}

\author{Chen Ma}
\authornote{Corresponding author}
\affiliation{%
  \institution{City University of Hong Kong}
  \country{Hong Kong SAR}}
\email{chenma@cityu.edu.hk}

\renewcommand{\shortauthors}{Bowei He et al.}


\begin{abstract}
Personalized item ranking has been a crucial component contributing to the performance of recommender systems. As a representative approach, pairwise ranking directly optimizes the ranking with user implicit feedback by constructing (\textit{user}, \textit{positive item}, \textit{negative item}) triplets. Several recent works have noticed that treating all triplets equally may hardly achieve the best effects. They assign different importance scores to negative items, user-item pairs, or triplets, respectively. However, almost all the generated importance scores are groundless and hard to interpret, thus far from trustworthy and transparent. To tackle these, we propose the \textit{Triplet Shapley}---a Shapely value-based method to measure the triplet importance in an interpretable manner. Due to the huge number of triplets, we transform the original Shapley value calculation to the Monte Carlo (MC) approximation, where the guarantee for the approximation unbiasedness is also provided. To stabilize the MC approximation, we adopt a control covariates-based method. Finally, we utilize the triplet Shapley value to guide the resampling of important triplets for benefiting the model learning. Extensive experiments are conducted on six public datasets involving classical matrix factorization- and graph neural network-based recommendation models. Empirical results and subsequent analysis show that our model consistently outperforms the state-of-the-art methods.

\end{abstract}





\maketitle

\section{Introduction}

Personalized ranking is widely regarded as the essence of recommender systems (RS)~\citep{bpr2008}. It ranks the recommended items according to user preferences to make the most relevant items appear at the top of the recommendation list, thereby improving user satisfaction and the benefits of service providers. 
To enable personalized ranking, implicit feedback (e.g., check-ins and clicks) is more broadly utilized than explicit feedback (e.g., ratings), due to its accessibility. Hereby, many approaches have been proposed using implicit feedback. 
Among them, Bayesian Personalized Ranking (BPR)~\citep{bpr2008} is one of the most representative paradigms that can be incorporated in many different types of methods like matrix factorization (MF)-based models~\citep{koren2009matrix, he2017neural, arora2019implicit}, and graph neural network (GNN)-based models~\citep{wang2019neural, he2020lightgcn, sun2019multi, max2017gcn, max2017gcmc, he2023dynamically}. In detail, BPR constructs (\textit{user}, \textit{positive item}, \textit{negative item}) triplets and optimizes the pairwise ranking score between positive and negative (randomly sampled) items for each triplet. Indeed, the quality of constructed triplets largely affects the ranking performance, especially when negative items are randomly sampled from non-interacted items. Because a user may have different preference levels for positive items, and it is hard to tell whether the randomly sampled items are truly negative and how negative they are. Therefore, it is highly necessary to further distinguish the importance of such triplets. 

\begin{figure*}[ht]
    \centering
    \includegraphics[width=0.95\textwidth]{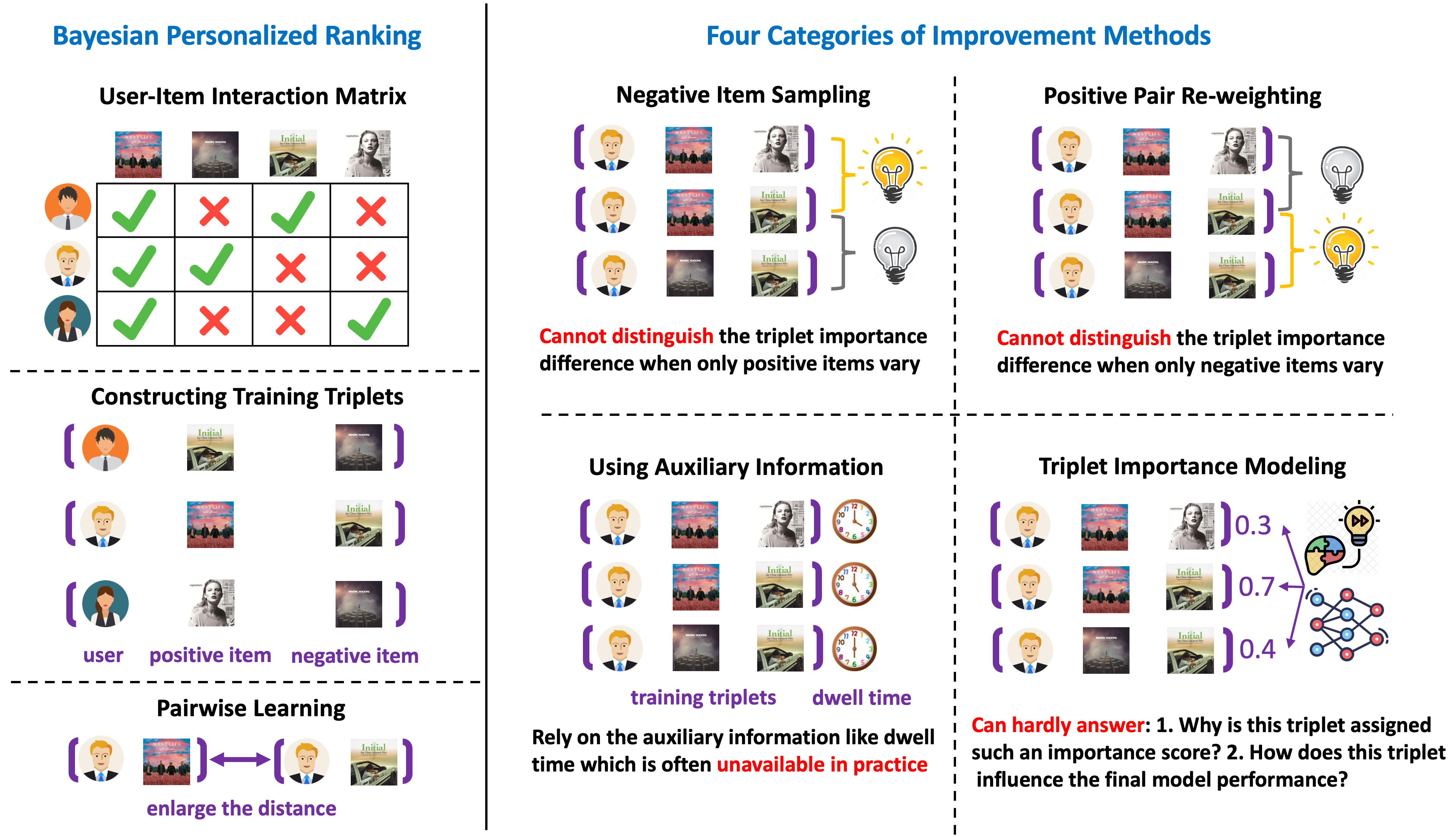}
    \caption{The illustration of vanilla BPR and corresponding four categories of improvement methods.
    }
    \Description{The illustration of vanilla BPR and corresponding four categories of improvement methods.}
    \label{fig: intro}
    \vspace{-0.3cm}
\end{figure*}

To discern the triplet importance, four main categories of methods have been proposed: \textit{negative item sampling}, \textit{positive pair re-weighting}, \textit{using auxiliary information}, and \textit{triplet importance modeling}. Among them, \textit{negative item sampling} techniques~\citep{gantner2012personalized, rendle2014improving, lian2020personalized} prioritize items differently during the training process, based on criteria such as item popularity or user engagement. While this approach can enhance the model's focus on more influential negative items, it can induce biases by over-representing popular items and under-representing less common but potentially significant items. \textit{Positive pair re-weighting} methods~\citep{wang2021denoising} adjust the influence of user-item interactions within the model by assigning different weights to them. The approaches in such two categories have recognized the necessity of applying different weights to positive or negative items. However, they only focus on the part of the triplet and neglect the whole, as illustrated in Figure~\ref{fig: intro}. For the methods of \textit{using auxiliary information} ~\citep{kim2014modeling, yi2014beyond, lu2018between, xie2022reweighting}, they rely on the side information like dwell time that is often unavailable in practical settings. In addition to these three types of methods, \textit{triplet importance modeling}~\citep{wu2022adapting} advances one step further and takes the whole triplet into consideration. By dispatching triplet-level importance, this method can distinguish the contribution of each triplet in the training. However, these works either use a heuristic approach or directly utilize a black-box neural network module ~\citep{wu2022adapting} to generate the triplet importance, where such importance scores are groundless and hard to interpret. The lack of interpretability will make both the importance score and the recommendation model less trustworthy, and even less fair due to the potential unknown bias introduced in importance generation.
Besides, they fail to bring insights on how to adjust the sampling strategy, which wastes massive high-importance triplets that can be easily accessed with resampling, thus restricting the further improvement of model performance.


To overcome aforementioned drawbacks, we propose a more interpretable approach to generate the importance of each triplet and further resample the important triplets to enhance the model performance, namely \textbf{I}nterpretable \textbf{T}riplet \textbf{I}mportance for \textbf{P}ersonalized \textbf{R}anking (ITIPR). First, based on previous works of Shapley values~\citep{datashapley, neuronshapley, disdatavalue}, we define the triplet Shapley to value the importance of each triplet, which discern their marginal contributions to the personalized ranking performance over all possible triplet subsets, thus achieving significant advantages in two aspects of interpretability and equitability. Second, we design a gradient descent-based Monte Carlo approximation method to compute triplet Shapley, whose theoretical guarantee of unbiasedness is also provided. To stabilize the triplet Shapley approximation, we construct a correlated variable and use the control variates method to help reduce the variance of the Monte Carlo estimator. Furthermore, we propose an importance-aware triplet resampling method for supplementing high-importance triplets to previously constructed ones. Here, we adopt a neural network module to learn to predict triplet Shapley for triplets to be sampled, thus facilitating the resampling process. 

To summarize, the main contributions are as follows:
\begin{itemize}[leftmargin=*]
    \item To achieve more interpretable and equitable triplet importance, we define the triplet Shapley as the importance score and design the Monte Carlo approximation method to estimate the Shapley value with unbiasedness guarantee. To alleviate the uncertainty of this estimator, we provide a control variates approach which only requires limited extra computation efforts.
    
    \item We propose an importance-aware resampling method to collect more high-importance triplets into the original triplet set, contributing to the model training.
    
    \item To demonstrate the effectiveness and applicability of our method, extensive experiments are conducted on six public datasets with two matrix factorization-based and two graph network network-based backbone models, whose results show ITIPR can outperform the state-of-the-art baselines.
    
\end{itemize}

\section{Related Work}
\textbf{Implicit Feedback in Recommendation}. Due to the easier accessibility than explicit feedback like rating scores\citep{salakhutdinov2007restricted, sarwar2001item} in most recommendation scenarios, the implicit feedback like historical clicking or browsing behaviors has become the main data source in personalized recommender system. Actually, there have been many recommendation models developed to utilize the implicit feedback for conducting personalized ranking, from the conventional MF~\citep{koren2009matrix} and KNN~\citep{deshpande2004item} to the recent deep neural network-based NeuMF~\citep{he2017neural}, NGCF~\citep{wang2019neural}, and LightGCN~\citep{he2020lightgcn}. BPR~\citep{bpr2008} first notices that few of previous methods are directly optimized for the personalized ranking and proposes to address this challenge by constructing (\textit{user}, \textit{positive item}, \textit{negative item}) triplets and pairwise training. In detail, BPR proposes the maximum posterior estimator derived from the Bayesian analysis to the problem. Based on this, some improvement schemes have also been proposed, like \textit{negative item sampling}~\citep{gantner2012personalized, rendle2014improving, lian2020personalized}, \textit{positive pair re-weighting}~\citep{wang2021denoising}, \textit{using auxiliary information}~\citep{kim2014modeling, yi2014beyond, lu2018between, xie2022reweighting, wen2019leveraging}. Futhermore, many real-world observations and empirical results have demonstrated that different triplets contribute differently to the model training performance. To prevent the performance loss due to treating all triplets equally, TIL~\citep{wu2022adapting} leverages a neural network to learn the triplet importance end-to-end. However, previous methods either use a heuristic approach or an unexplainable blackbox model to generate the triplet importance. It is hard to understand why assigning such triplet importance scores to corresponding triplets and how these triplet importance scores are connected with the final recommendation model performance. 

\textbf{Data Valuation}. Data valuation aims at quantifying the contribution of data in the algorithmic predictions and decisions, considering the it has been the main power source of data science. Leave-one-out (LOO) test~\citep{Cook00, hossain1989detection} is a kind of traditional data valuation approach. It is often approximated by the leverage or influence score~\citep{cook1982residuals} which measures the model output change when removing the studied sample. However, LOO exhibits the drawback of lacking equitability and performs badly in many real applications. To address such issues, data Shapley originating from cooperative game theory research is proposed\citep{JiaDWHGLZSS19, datashapley, disdatavalue, neuronshapley, betashapley,yan2021if, schoch2022cs}. This is an intuitive and useful tool to value the contribution of each datum in a dataset and has attacted wide attention recently. Nevertheless, directly computing the Shapley value according to the original definition is pretty costly and can hardly be promoted to the real application. To address this issue, some acceleration schemes have also been proposed\citep{disshapley, jethani2021fastshap}. 
However, the existing data Shapley methods are mainly designed for traditional regression and classification tasks in computer vision, lacking the discussion in the personalized ranking tasks which is the focus of this paper.

\section{Preliminaries}
In this section, we first formulate the top-$K$ recommendation task. Then, we review the Bayesian Personalized Ranking paradigm which is the foundation of our method.

\subsection{Problem Formulation}
The recommendation task in this paper leverages the user implicit feedback as input. Let $\mathcal{U}$ and $\mathcal{V}$ denote the set of all users and items in the system, respectively. $\mathbf{Y} \in \mathbb{R}^{|\mathcal{U}| \times |\mathcal{V}|}$ is the binary implicit feedback rating matrix. For each user $u \in \mathcal{U}$, the user preference data is represented by the set of items in his/her interaction history as $\mathcal{I}_u = \left\{i \in \mathcal{V} | \mathbf{Y}_{u,i}=1\right\}$. The top-$K$ recommendation task is then formulated: for each user $u \in \mathcal{U}$, given the training item set $\mathcal{S}_u$ and the non-empty test item set $\mathcal{T}_u$ (requiring that $\mathcal{S}_u \cup \mathcal{T}_u = \mathcal{I}_u$ and $\mathcal{S}_u \cap \mathcal{T}_u = \emptyset$), the model needs to recommend an ordered item set $\mathcal{X}_u$ such that $|\mathcal{X}_u| = K$ and $\mathcal{X}_u \cap \mathcal{S}_u = \emptyset$. Then, the recommendation quality is evaluated with some matching scores between $\mathcal{X}_u$ and $\mathcal{T}_u$, like Recall@$K$ and NDCG@$K$. 

\subsection{Bayesian Personalized Ranking}
Bayesian Personalized Ranking (BPR) has been widely integrated with many recommendation models and achieved great success in many real-world production scenarios~\citep{koren2009matrix, he2017neural, wang2019neural, he2020lightgcn, he2023dynamic}. The BPR method mainly consists of two parts: triplet construction and pairwise training. The details of them are as follows.

\textbf{Triplet Construction}. Based on the assumption that the users prefer the interacted (positive) items over the uninteracted (negative) ones, the triplet construction space $D_S = \mathcal{U} \times \mathcal{V} \times \mathcal{V}$ is first formalized by:
\begin{equation}
    D_S := \{ (u,i,j) | i \in \mathcal{I}_u, j \in \mathcal{V} \textbackslash \mathcal{I}_u \},
\end{equation}
where for triplet $(u,i,j) \in D_S$, $i$ is the positive item, and $j$ is the negative item. However, directly traversing all the triplets in $D_S$ for recommendation modeling learning will bring huge computation costs and poor convergence, which hinders effective training. Therefore, BPR chooses the triplets uniformly from the triplet construction space $D_S$. In detail, for each user and positive item pair $(u,i)$, BPR uniformly samples a constant number (usually no more than ten) of negative items $j$, and thus obtains a triplet set $D^T$ for further training recommendation models.

\textbf{Pairwise Training}. Based on the Bayesian formulation of learning the correct personalized ranking for all items, BPR proposes the maximum posterior estimator to derive the generic optimization criterion (loss function) for personalized ranking. Specifically, BPR contrasts the user-positive item pair $(u,i)$ with user-negative item pair $(u,j)$ in each triplet $(u,i,j) \in D^T$ to guide the model capturing the user $u$'s preference:
\begin{equation}
    \mathcal{L}_{BPR} (u,i,j;\boldsymbol{\theta}) = - ln \ \sigma(\hat{y}_{ui}(\boldsymbol{\theta}) - \hat{y}_{uj}(\boldsymbol{\theta})),
\end{equation}
where $\boldsymbol{\theta}$ indicates the parameters of the recommendation model, like MF-based (MF~\citep{koren2009matrix}, NeuMF~\citep{he2017neural}) and GNN-based ones (NGCF~\citep{wang2019neural}, LightGCN~\citep{he2020lightgcn}). $\hat{y}$ is an arbitrary real value representing the model's prediction of the user preference. $\sigma(\cdot)$ is the Sigmoid function.

\section{Methodology}
In this section, we first follow the \textit{triplet importance model} approach and introduce the triplet importance-based weighted pairwise learning. Then, we propose the triplet Shapley as the interpretable and equitable importance to tackle the transparency and fairness issues of the existing methods. It also serves as the more discriminative triplet weight to facilitate pairwise learning. Next, we re-examine the triplet Shapley formulation from the joining process perspective and utilize the descent-based Monte Carlo method to approximate the triplet Shapley. Besides, we provide a control covariates scheme to stabilize the triplet Shapley estimation. Finally, we build a triplet importance prediction model to guide the triplet resampling, so as to adaptively augment the original constructed triplets.

\subsection{Triplet Importance for Pairwise Learning}
Although Bayesian personalized ranking has been successfully integrated into many recommendation models, it equally treats all triplets which may not capture the fine-grained user preference, since users have different preference levels on different items. Moreover, a randomly sampled item does not necessarily mean that the user dislikes it, especially for those similar to the positive items. These motivate us to assign importance scores to different triplets to facilitate the personalized ranking. To incur the triplet importance into the original BPR loss, we adopt the formulation from the previous work~\citep{wu2022adapting} as follows,
\begin{equation}
    \mathcal{L}_{TI-BPR} (u,i,j;\boldsymbol{\theta}) = - W^{TI}_{(u,i,j)} \cdot ln \ \sigma(\hat{y}_{ui}(\boldsymbol{\theta}) - \hat{y}_{uj}(\boldsymbol{\theta})),
\label{equ:wpl}
\end{equation}
where $W^{TI}_{(u,i,j)}$ is the triplet importance of the triplet $(u,i,j)$ for the BPR loss.

\subsection{Triplet Shapley as Interpretable Importance}
\label{sec:triplet shapley}

Although previous approaches~\citep{gantner2012personalized, rendle2014improving, lian2020personalized, wang2021denoising, wu2022adapting} have been proven to enhance Bayesian pairwise user preference learning, few of them pay attention to the interpretability and equitability of the modeled triplet importance. 
Moreover, due to the black-box property of most previous methods, their optimality in distinguishing different triplets can hardly be guaranteed. On the other hand, the Shapley value, which originates from the game theory and data valuation, has demonstrated the capability of identifying the data importance in an explicit way~\citep{datashapley}. 
The advantage of employing the Shapley value in the domain of personalized ranking lies in its ability to fairly distribute the total gain of a collaborative effort among all contributing triplets. By integrating Shapley values, we aim to allocate credit among constructed triplets, thereby providing insights into which triplets mostly influence the ranking performance. This approach foregrounds equitable valuation and significantly heightens interpretability by providing a detailed understanding of data influence, that is not solely confined to traditional prediction tasks~\citep{datashapley, betashapley, jethani2021fastshap}. To adapt the Shapley value into the personalized ranking, we define the triplet Shapley as interpretable importance:

\begin{definition}{\textbf{Triplet Shapley}.} The triplet Shapley for a triplet $(u, i, j) \in D^{T}$  is defined as:
\begin{equation}
    V^{TS}_{(u,i,j)} = C \underset{S \subset \{(u,i,j)\}^{'}}{\sum} \frac{A(S \cup \left\{(u,i,j)\}\right) - A(S)}{\binom{|D^{T}|-1}{|S|}},
\label{equ:definition}
\end{equation}
\end{definition}
\noindent where $\{(u,i,j)\}^{'}$ indicates the complement set $D^{T} \backslash \{(u,i,j)\}$. $A(S)$ denotes the performance on validation data of the recommendation model trained with triplet subset $S$. Here, the $C$ is a constant to be determined. Based on the definition, we can derive the interpretability and equitability for more transparent and fair triplet importance measurement. The corresponding details are described below.

\textbf{Interpretability:} The triplet Shapley is defined based on its direct influence on the final model performance. It explicitly reflects the triplet's contribution to the model performance. Furthermore, from the perspective of information theory, triplet Shapley quantifies how much information is gained for model training from including each triplet into the training set $D^{T}$. For an \textit{important} triplet with a high triplet Shapley value, removing it from the training set will lead to obvious recommendation accuracy degradation.

\textbf{Equitability}: The triplet Shapley exhibits following properties that guarantee its equitability: 1) \textit{Inclusiveness} property: considering the triplet's marginal contribution under each possible subset scenario; 2) \textit{Equal value condition} property: assigning identical Shapley values to triplets that contribute equally to the model’s predictions, regardless of the subset they're in. Such properties can be obtained by following the traditional Shapley value analysis~\citep{dubey1975uniqueness, roth1988shapley} and regarding each triplet here as a player in the cooperative game.



\subsection{Approximating Triplet Shapley}
\label{sec:approx}

\subsubsection{\textbf{Identical Transformation to Triplet Shapley}}
\label{sec: identical transform}
For more effectively computing the triplet Shapley defined in Eq.~\ref{equ:definition}, we need to first conduct an identical transformation to it.
\begin{theorem}{\textbf{Triplet Shapley from Joining Process Perspective.}}
Denote by $\Pi^{T}$ the set of all possible permutations of the triplets in $D^{T}$, each of which representing a distinct joining order. Moreover, let $P^{\pi}_{(u,i,j)}$ denote the set of triplets that precede $(u,i,j)$ in the permutation $\pi \in \Pi^{T}$. Then, the triplet Shapley of (u,i,j) defined in Eq.~\ref{equ:definition} can be calculated as:
\begin{equation}
    V^{TS}_{(u,i,j)} = \frac{1}{|D^{T}|!} \underset{\pi \in \Pi^{T}}{\sum} A(P^{\pi}_{(u,i,j)} \cup \left\{(u,i,j)\}\right) - A(P^{\pi}_{(u,i,j)}).
\label{equ:the1}
\end{equation}
\label{theo1}
\end{theorem}
\vspace{-3mm}
This theorem can be proved by setting $C = \frac{1}{|D^{T}|}$ in Eq.~\ref{equ:definition}.

\subsubsection{\textbf{Monte Carlo Approximation}}
Monte Carlo approximation has been widely used in mathematics and physics~\citep{karp1989monte}. This kind of approach relies on the repeated random sampling to obtain the numerical results for some super complex or difficult problems. In this work, we also utilize the Monte Carlo approximation technique to estimate the triplet Shapley via randomly sampling possible permutations based on the above transformation in Sec~\ref{sec: identical transform}. To elaborate on the theoretical rationality of Monte Carlo approximation, we provide the following theorem on the unbiasedness of Monte Carlo triplet Shapley approximation:
\begin{theorem}{\textbf{Unbiasedness of Monte Carlo Approximation.}}
For $V^{TS, \pi}_{(u,i,j)}$, its average value $\overline{V^{TS, \pi}_{(u,i,j)}}$ over the different permutations $\pi_k$ randomly sampled from the uniform distribution $U(\pi)$ is the unbiased estimator of the corresponding triplet Shapley $V^{TS}_{(u,i,j)}$. 
\label{theo2}
\end{theorem}

\subsubsection{\textbf{Efficient Triplet Shapley Approximation}} However, directly applying Monte Carlo approximation will still face the challenges brought by the slow convergence of recommendation model training and the intrinsic noise as the permutation stretches. To address such challenges, we specially design the gradient descent-based truncated Monte Carlo approximation for triplet Shapley, illustrated in the Algorithm~\ref{algorithm:mca}. The outer loop still follows the standard Monte Carlo approximation procedure. However, in the inner loop, we first use the one-epoch training with the studied triplet $\pi_k[s]$ (Line 11) instead of the conventional recommendation model training, which conducts the multi-epoch gradient descent with the same data and leads to the slow convergence. To achieve this goal and guarantee that the model has approximately converged, we need to adopt a relatively bigger learning rate $\alpha$ than that in the traditional multi-epoch training. In fact, this point has been proved to be feasible in ~\citep{zhang2022towards}. Meanwhile, considering that the model performance change $a^k_s - a^k_{s-1}$ becomes less and less obvious which can be easily drowned out by the noise as the the permutation stretches from $1$ to $|D^{T}|$ ($s \to |D^{T}|$), we design a truncation operation to the permutation scanning (Line 8, 9). In detail, we set a \textit{tolerance} threshold and truncate the permutation scanning when the $a^k_s$ has been closed enough to the model performance with the whole triplet set $A(D^{T})$. Then, the marginal contribution $a^k_s - a^k_{s-1}$ of all the rest triplets that follow the $s$-th position in this permutation will be assigned to zero, in such a way to effectively avoid the huge amounts of the uninformative calculations.

\begin{algorithm}
    \caption{Gradient Descent-Based Truncated Monte Carlo Approximation for Triplet Shapley.}
    \label{algorithm:mca}
    \KwIn{Previously collected triplet set $D^{T}$, parameterized recommendation model $M(\boldsymbol{\theta})$, vanilla BPR loss function $\mathcal{L}_{BPR}(\cdot;\boldsymbol{\theta})$, recommendation accuracy $Acc(\boldsymbol{\theta})$ (instantiation of $A(\cdot)$ with $\boldsymbol{\theta}$ as input).}
    \KwOut{Triplet Shapley of each triplet $(u,i,j)$ in $D^{T}$.}
    Initialize $V^{TS}_{u,i,j}=0$ for each triplet in $D^{T}$ and $k=0$.\\
    \While{Convergence criteria not met}{
      $k = k+1$;\\
      Randomly sample a permutation $\pi_k$ from the $U(\pi)$;\\
      Randomly initialize parameters of model $M(\boldsymbol{\theta})$ as $\boldsymbol{\theta}^k_0$;\\
      Initialize $a^k_0 = Acc(\boldsymbol{\theta}^k_0)$;\\
      \For{$s=1,...,|D^{T}|$}{
        \eIf{$|Acc(D^{T}) - a^k_{s-1}| < tolerance$}{
            $a^k_{s} = a^k_{s-1}$;
        }{
            $\boldsymbol{\theta}^k_s = \boldsymbol{\theta}^k_{s-1} - \alpha \nabla_{\boldsymbol{\theta}} \mathcal{L}_{BPR}(\pi_k[s];\boldsymbol{\theta}^k_{s-1}) $;\\
            $a^k_s = Acc(\boldsymbol{\theta}^k_s)$;\\
            $V^{TS}_{\pi_k[s]} = \frac{(k-1)}{k}V^{TS}_{\pi_{k-1}[s]} + \frac{1}{k}(a^k_{s} - a^k_{s-1})  $;\\
        }      
      }
      
    }
\end{algorithm}
\vspace{-1mm}
\subsubsection{\textbf{Stabilizing Triplet Shapley Approximation}} Though the use of Monte Carlo Sampling Approximation brings significant computational cost reduction, an accompanying issue of value uncertainty has also been raised. The direct consequence is that the variance of each approximated triplet Shapley value may obscure its corresponding true value. Thus, it becomes infeasible to utilize these estimates to effectively differentiate the importance of different triplets. To stabilize the triplet Shapley approximation and provide reliable triplet importance scores, we adopt the control variate method, which has been explored in machine learning and statistics~\citep{karimireddy2020scaffold, mohamed2020monte, horvath2023stochastic, sun2023vector} to reduce the variance of Monte Carlo estimators. In specific, it exploits information about the errors in known estimators to reduce the error of an unknown estimator. 

Suppose $\widehat{V^{TS}_{(u,i,j)}}$ and  $\widehat{V^{TS, *}_{(u,i,j)}}$ are unbiased estimators of $V^{TS}_{(u,i,j)}$ and $V^{TS, *}_{(u,i,j)}$, where $V^{TS, *}_{(u,i,j)}$ is known and correlated with $V^{TS}_{(u,i,j)}$. Thus, the control variates-based estimator can be defined as:
\begin{equation}
    \widetilde{V^{TS}_{(u,i,j)}} := \widehat{V^{TS}_{(u,i,j)}} - c(\widehat{V^{TS, *}_{(u,i,j)}} - V^{TS, *}_{(u,i,j)}).
\label{equ:cv}
\end{equation}
Here, $c$ is a constant. First, we can derive that Eq.~\ref{equ:cv} is the unbiased estimator of $V^{TS}_{(u,i,j)}$:
\begin{equation}
\begin{aligned}
    E(\widetilde{V^{TS}_{(u,i,j)}}) &= E(\widehat{V^{TS}_{(u,i,j)}}) - c E(\widehat{V^{TS, *}_{(u,i,j)}} - V^{TS, *}_{(u,i,j)}) \\
    & =E(\widehat{V^{TS}_{(u,i,j)}}) - c (E(\widehat{V^{TS, *}_{(u,i,j)}}) - E(V^{TS, *}_{(u,i,j)})) \\
    & = E(\widehat{V^{TS}_{(u,i,j)}}) = E(V^{TS}_{(u,i,j)}).
\end{aligned}
\end{equation}
Then, we propose the following lemma which can facilitate the selection of the optimal $c$ to minimize the variance of Eq.~\ref{equ:cv}.
\begin{lemma}{\textbf{Variance Reduction with Control Variates.}} 
    Suppose $\widehat{V^{TS}_{(u,i,j)}}$ and $\widehat{V^{TS, *}_{(u,i,j)}}$ are unbiased estimators and their Pearson's correlation coefficient $\rho(\widehat{V^{TS}_{(u,i,j)}}, \widehat{V^{TS, *}_{(u,i,j)}}) > 0$. Taking $c^* = \frac{Cov(\widehat{V^{TS}_{(u,i,j)}}, \widehat{V^{TS, *}_{(u,i,j)}})}{Var(\widehat{V^{TS, *}_{(u,i,j)}})}$, the control variates estimator $\widetilde{V^{TS}_{(u,i,j)}}$ has the minimal variance $Var(\widetilde{V^{TS}_{(u,i,j)}}) = (1-\rho^2(\widehat{V^{TS}_{(u,i,j)}}, \widehat{V^{TS, *}_{(u,i,j)}})) Var(\widehat{V^{TS}_{(u,i,j)}})$.
\label{lemma1}
\end{lemma}
Next, recall the gradient descent process in Algorithm~\ref{algorithm:mca} where we compute the gradient with the triplets one by one along the order of the permutation $\pi$. This kind of method, noted as $\Omega$, can help distinguish the contribution of each triplet clearly but needs to traverse all possible permutations to achieve the accurate triplet Shapley expectation. To construct a \textit{known} $V^{TS, *}_{(u,i,j)}$ whose value can be accurately obtained rather than being estimated with Monte Carlo sampling, we propose a modified version of gradient descent $\Omega^*$. In detail, for a triplet $(u,i,j)$ in $\pi$, we first initialize the model parameters $\boldsymbol{\theta} (\emptyset)$ from scratch and compute the gradient with all the triplets in $\{(u,i,j)\}^{'}$ concurrently. Then, we update the parameters to $\boldsymbol{\theta} (P^{\pi}_{(u,i,j)})$ with adjusted-step gradient descent:
\begin{equation}
    \boldsymbol{\theta} (P^{\pi}_{(u,i,j)}) = \boldsymbol{\theta} (\emptyset) - \alpha \frac{|P^{\pi}_{(u,i,j)}|}{|D^{T}|-1} \nabla_{\boldsymbol{\theta}} \mathcal{L}_{BPR}(\{(u,i,j)\}^{'};\boldsymbol{\theta} (\emptyset)).
\end{equation}
Finally, we compute the gradient with the single triplet $(u,i,j)$ and update the model parameter to $\boldsymbol{\theta} (P^{\pi}_{(u,i,j)}\cup \left\{(u,i,j)\}\right)$. We use the triplet Shapley computed with this modified gradient descent algorithm $\Omega^{*}$ as $V^{TS, *}_{(u,i,j)}$. In following theorem, we show that its accurate value can be obtained within acceptable computation budget.

\begin{theorem}{\textbf{Triplet Shapley with Modified Gradient Descent.}} 
Let $A^{*}(P^{\pi}_{(u,i,j)})$ and $A^{*}(P^{\pi}_{(u,i,j)} \cup \left\{(u,i,j)\}\right)$ denote the performance on validation data of model parameterized with the above $\boldsymbol{\theta} (P^{\pi}_{(u,i,j)}\cup \left\{(u,i,j)\}\right)$ and $\boldsymbol{\theta} (P^{\pi}_{(u,i,j)})$, respectively. Following Eq.~\ref{equ:the1}, we have:
\begin{equation}
    V^{TS, *}_{(u,i,j)} = \frac{1}{|D^{T}|!} \underset{\pi \in \Pi^{T}}{\sum} A^{*}(P^{\pi}_{(u,i,j)} \cup \left\{(u,i,j)\}\right) - A^{*}(P^{\pi}_{(u,i,j)}).
\label{equ:the31}
\end{equation}
Thus, with only $|D^{T}|$ specific permutations (much less that $|D^{T}|!$), the accurate value of $ V^{TS, *}_{(u,i,j)}$ for all triplets can be obtained with:
\begin{equation}
V^{TS, *}_{(u,i,j)} = \frac{1}{|D^{T}|} \underset{k=1}{\sum^{|D^{T}|}} A^{*}(P^{\pi_k}_{(u,i,j)} \cup \left\{(u,i,j)\}\right) - A^{*}(P^{\pi_k}_{(u,i,j)}).
\label{equ:the32}
\end{equation} 
Such $|D^{T}|$ permutations need to meet the requirement that for each triplet $(u,i,j)$, its position in the permutation should be totally different among all $|D^{T}|$ permutations. To obtain these permutations, we first initialize a permutation containing all triplets in $D^{T}$ randomly. Then, we dequeue the triplet at its last position and enqueue it into the first position to get a new permutation. Following this process, we can obtain $|D^{T}|$ different permutations that meet the above requirement.
\label{theo3}
\end{theorem}


Finally, to determine the value of $c$ in Eq.~\ref{equ:cv}, we still need to compute the $Var(\widehat{V^{TS, *}_{(u,i,j)}})$ and $Cov(\widehat{V^{TS}_{(u,i,j)}}, \widehat{V^{TS, *}_{(u,i,j)}})$ according to Lemma~\ref{lemma1}. Considering we have a certain degree of tolerance to $c$'s selection error, we directly estimate their values empirically. In each iteration of the Algorithm~\ref{algorithm:mca}, we can obtain an observed value $V^{TS, \pi}_{(u,i,j)}$ and $V^{TS, *, \pi}_{(u,i,j)}$ for $V^{TS}_{(u,i,j)}$ and $V^{TS, *}_{(u,i,j)}$, respectively. Thus, 
\begin{equation}
\begin{aligned}
    Var(\widehat{V^{TS, *}_{(u,i,j)}}) &= \frac{1}{n} Var(V^{TS, *, \pi}_{(u,i,j)}), \\
    Cov(\widehat{V^{TS}_{(u,i,j)}}, \widehat{V^{TS, *}_{(u,i,j)}}) &= \frac{1}{n} Cov(V^{TS, \pi}_{(u,i,j)}, V^{TS, *, \pi}_{(u,i,j)}),
\end{aligned}
\end{equation}
where $n$ is the whole number of iterations.

\subsection{Importance-aware Triplet Resampling}
Previous works on distinguishing triplet-level importance~\citep{wu2022adapting} just focus on how to make use of the triplet set $D^{T}$ constructed at the beginning well, while ignoring exploiting the data potential in the whole triplet construction space ($|\mathcal{U}|\times|\mathcal{V}|^2$). Actually, other potential triplets not sampled into the $D^{T}$ at first can also bring informative contributions to our recommendation model training. Besides, the triplet Shapley values for elements in $D^{T}$ approximated in Sec.~\ref{sec:approx} can exactly inform us which kind of triplets are high-importance and which are low-importance. To utilize this information well and effectively estimate the triplet Shapley for other potential triplets not in $D^{T}$, we first fit a neural network model to learn predicting their triplet Shapley values, which is named as \textbf{T}riplet \textbf{I}mportance \textbf{P}rediction (TIP) model. Without sophisticated design, we just use a simple two-layer MLP structure here:
\begin{equation}
\begin{aligned}
    \mathbf{h}_{(u,i,j)} &= ReLU(\mathbf{W}_1 \cdot [\mathbf{p}_u; \mathbf{q}_i; \mathbf{q}_j] + \mathbf{b}_1), \\
    \widehat{V^{TS}_{(u,i,j)}} &= Tanh(\mathbf{W}_2 \cdot \mathbf{h}_{(u,i,j)} + \mathbf{b}_2).
\label{equ:impred}
\end{aligned}
\end{equation}
Next, to fully unleash the potential of the whole triplet construction space $D_S$, we rely on such kind of triplet importance information to guide the resampling process. In detail, when resampling triplets, given user $u$ and positive item $i$, we model the sampling probability of the unsampled negative item $j ((u,i,j) \notin D^{T})$ as follows:
\begin{equation}
\begin{aligned}
    p(j|u, i) &= \frac{e^{\mathbf{TIP}(u,i,j)}}{\underset{\substack{j' \in \mathcal{V}\backslash\mathcal{I}_u \\ (u,i,j') \notin D^{T}}}{\sum} e^{\mathbf{TIP}(u,i,j')}}.
\label{equ:resample}
\end{aligned}
\end{equation}

\subsection{Overall Learning Procedure}
To better illustrate the workflow and facilitate the reproducibility, we summarize the overall learning procedure of our proposed method in Algorithm~\ref{algorithm:olp}. It can be generally divided into three parts: \textit{triplet importance approximation}, \textit{importance-aware triplet resampling}, and \textit{triplet importance-weighted pairwise learning}. Note that before training recommendation model $M(\boldsymbol{\theta})$ with the augmented triplet set $D^T_{aug}$, we need to reestimate the triplet Shapley $V^{TS}_{(u,i,j)}$ for each triplet in $D^T_{aug}$, in this way to take the influence brought by the newly resampled triplets into consideration. In addition, we perform a normalization operation on the reestimated $V^{TS}_{(u,i,j)}$ to rescale them into a reasonable value range, which are then taken as weights for corresponding BPR losses (Eq.~\ref{equ:wpl}) during training.
\begin{algorithm}
    \caption{Overall Learning Procedure of Interpretable Triplet Importance for Personalized Ranking.}
    \label{algorithm:olp}
    \KwIn{User set $\mathcal{U}$, item set $\mathcal{V}$, user-item interaction set $\mathcal{I}$, previously collected triplet set $D^{T}$}
    \KwOut{Trained recommendation model $M(\boldsymbol{\theta})$}
     $/*$ \textit{Triplet Importance Approximation} $*/$ \\
     Estimate the triplet Shapley $V^{TS}_{(u,i,j)}$ for each $(u,i,j)$ in $D^T$ with Algorithm~\ref{algorithm:mca} and Eq.~\ref{equ:cv}; \\
     $/*$ \textit{Importance-aware Triplet Resampling} $*/$ \\
     Learn the TIP model with the triplet Shapley $V^{TS}_{(u,i,j)}$ in $D^T$; \\ 
     Resample new triplets into $D^T_{new}$ with Eq.~\ref{equ:resample}; \\
     Combine $D^T_{new}$ and original $D^{T}$ to obtain augmented $D^T_{aug}$; \\
     $/*$ \textit{Triplet Importance-weighted Pairwise Learning} $*/$ \\
     Reestimate the the triplet Shapley $V^{TS}_{(u,i,j)}$ for each $(u,i,j)$ in $D^T_{aug}$ with Algorithm~\ref{algorithm:mca} and Eq.~\ref{equ:cv}; \\
     Train the recommendation model $M(\boldsymbol{\theta})$ from scratch with Eq.~\ref{equ:wpl} as loss function and $D^T_{aug}$;
\end{algorithm}

\section{Experiments}
In this section, we conduct experiments on several public datasets to demonstrate the effectiveness, applicability, interpretability, and time efficiency of our method. In the experiments, we mainly focus on answering the following questions:

\begin{itemize}[leftmargin=*]

\item \textbf{RQ1:} Whether our method can achieve better recommendation performance than other baselines utilizing implicit feedback?

\item \textbf{RQ2:} Whether the superiority of our method can be consistently performed when selecting different MF-based and  GNN-based recommendation models as the backbone?

\item \textbf{RQ3:} Does each component of our method contribute to the performance improvement (Ablation study)?

\item \textbf{RQ4:} Can our method provide a reasonable and intuitive explanation to the importance score assigned to each triplet?

\item \textbf{RQ5:} Whether our control covariate-based method can effectively stabilize the triplet Shapley approximation, thus leading to more reliable triplet Shapley values?

\item \textbf{RQ6:} Whether the time complexity of our method is still acceptable considering the substantial computation cost brought by the triplet Shapley? In other words, Can our approximation approach effectively reduce the training time cost?

\end{itemize}
\vspace{-2mm}

\begin{table*}[]
\caption{The performance comparison of all methods with different base models on six datasets in terms of \textit{R@20(Recall@20)} and \textit{N@20(NDCG@20)}. The best and second best results are marked in bold and underlined, respectively. $*$ indicates the improvements over baselines are statistically significant ($t$-test, $p$-value $\leq 0.01$).}
\vspace{-2mm}
\resizebox{0.97\textwidth}{!}{
\begin{tabular}{|c|cc|cc|cc|cc|cc|cc|}
\hline
\multirow{2}{*}{Methods} & \multicolumn{2}{c|}{Amazon-Books}       & \multicolumn{2}{c|}{Amazon-CDs}         & \multicolumn{2}{c|}{Gowalla}     & \multicolumn{2}{c|}{Yelp}        & \multicolumn{2}{c|}{ML-20M}   & \multicolumn{2}{c|}{ML-Latest}      \\ \cline{2-13}
                        & \multicolumn{1}{c|}{R@20} & N@20 & \multicolumn{1}{c|}{R@20} & N@20 & \multicolumn{1}{c|}{R@20} & N@20 & \multicolumn{1}{c|}{R@20} & N@20 & \multicolumn{1}{c|}{R@20} & N@20 & \multicolumn{1}{c|}{R@20} & N@20 \\ \hline
                        \multicolumn{13}{|c|}{MF} \\\hline
BPR                     & \multicolumn{1}{c|}{0.0209}     &  0.0141    & \multicolumn{1}{c|}{0.0999}     &  0.0581    & \multicolumn{1}{c|}{0.1219}     & 0.0776     & \multicolumn{1}{c|}{0.0564}     &  0.0315    & \multicolumn{1}{c|}{0.1763}     & 0.1327     & \multicolumn{1}{c|}{0.1473}     &   0.1116   \\ \hline
WBPR                    & \multicolumn{1}{c|}{0.0245}     &   0.0168   & \multicolumn{1}{c|}{0.1001}     &  0.0592    & \multicolumn{1}{c|}{0.1302}     &   0.0792   & \multicolumn{1}{c|}{0.0569}     & 0.0327     & \multicolumn{1}{c|}{0.1851}     & 0.1371     & \multicolumn{1}{c|}{0.1631}     &  0.1247    \\ \hline
AOBPR                   & \multicolumn{1}{c|}{0.0258}     &  0.0173    & \multicolumn{1}{c|}{0.1003}     & 0.0583     & \multicolumn{1}{c|}{0.1222}     &   0.0780   & \multicolumn{1}{c|}{0.0573}     &  0.0332    & \multicolumn{1}{c|}{0.1866}     & 0.1380     & \multicolumn{1}{c|}{0.1644}     &  0.1259    \\ \hline
PRIS                    & \multicolumn{1}{c|}{0.0337}     & 0.0235     & \multicolumn{1}{c|}{0.1032}     &  0.0604    & \multicolumn{1}{c|}{0.1488}     &  0.0949    & \multicolumn{1}{c|}{0.0745}     &  0.0414    & \multicolumn{1}{c|}{0.2342}     &  0.1709    & \multicolumn{1}{c|}{0.2111}     & 0.1562     \\ \hline
TCE                     & \multicolumn{1}{c|}{0.0293}     &  0.0190    & \multicolumn{1}{c|}{0.0890}     & 0.0503     & \multicolumn{1}{c|}{0.1305}     &  0.0806    & \multicolumn{1}{c|}{0.0647}     &  0.0376    & \multicolumn{1}{c|}{0.2038}     &  0.1554    & \multicolumn{1}{c|}{0.1985}     &  0.1483    \\ \hline
RCE                     & \multicolumn{1}{c|}{0.0286}     &  0.0185    & \multicolumn{1}{c|}{0.0970}     &  0.0569    & \multicolumn{1}{c|}{0.1283}     &  0.0782    & \multicolumn{1}{c|}{0.0676}     &  0.0384    & \multicolumn{1}{c|}{0.1975}     & 0.1513     & \multicolumn{1}{c|}{0.1952}     &  0.1475    \\ \hline
TCE-BPR                 & \multicolumn{1}{c|}{0.0379}     &  0.0257    & \multicolumn{1}{c|}{0.0903}     &  0.0536    & \multicolumn{1}{c|}{0.1394}     &  0.0853    & \multicolumn{1}{c|}{0.0662}     & 0.0380     & \multicolumn{1}{c|}{0.2124}     &  0.1610    & \multicolumn{1}{c|}{0.2093}     &  0.1547    \\ \hline
RCE-BPR                 & \multicolumn{1}{c|}{0.0364}     &  0.0242    & \multicolumn{1}{c|}{0.0971}     & 0.0573     & \multicolumn{1}{c|}{0.1382}     &  0.0825    & \multicolumn{1}{c|}{0.0697}     &  0.0395    & \multicolumn{1}{c|}{0.2086}     & 0.1676     & \multicolumn{1}{c|}{0.2054}     &  0.1516    \\ \hline
TIL-UI                  & \multicolumn{1}{c|}{0.0602}     & 0.0398     & \multicolumn{1}{c|}{0.1088}     & 0.0630     & \multicolumn{1}{c|}{0.1580}     & 0.0977     & \multicolumn{1}{c|}{0.0910}     & 0.0485    & \multicolumn{1}{c|}{0.2657}     &  0.1940    & \multicolumn{1}{c|}{0.2466}     & 0.1739     \\ \hline
TIL-MI                  & \multicolumn{1}{c|}{\underline{0.0614}}     & \underline{0.0407}     & \multicolumn{1}{c|}{\underline{0.1203}}     &  \underline{0.0711}    & \multicolumn{1}{c|}{\underline{0.1642}}     &  \underline{0.1009}    & \multicolumn{1}{c|}{\underline{0.0947}}     &  \underline{0.0503}    & \multicolumn{1}{c|}{\underline{0.2759}}     &   \underline{0.1988}   & \multicolumn{1}{c|}{\underline{0.2618}}     & \underline{0.1834}     \\ \hline
ITIPR                 & \multicolumn{1}{c|}{\textbf{\begin{tabular}[c]{@{}c@{}}0.0671*\\ ($\uparrow$ 9.28\%)\end{tabular}}}     &   \textbf{\begin{tabular}[c]{@{}c@{}}0.0459*\\ ($\uparrow$ 12.78\%)\end{tabular}}   & \multicolumn{1}{c|}{\textbf{\begin{tabular}[c]{@{}c@{}}0.1282*\\ ($\uparrow$ 6.57\%)\end{tabular}}}     & \textbf{\begin{tabular}[c]{@{}c@{}}0.0746*\\ ($\uparrow$ 4.92\%)\end{tabular}}     & \multicolumn{1}{c|}{\textbf{\begin{tabular}[c]{@{}c@{}}0.1786*\\ ($\uparrow$ 8.77\%)\end{tabular}}}     &   \textbf{\begin{tabular}[c]{@{}c@{}}0.1071*\\ ($\uparrow$ 6.14\%)\end{tabular}}   & \multicolumn{1}{c|}{\textbf{\begin{tabular}[c]{@{}c@{}}0.1032*\\ ($\uparrow$ 8.98\%)\end{tabular}}}     &   \textbf{\begin{tabular}[c]{@{}c@{}}0.0561*\\ ($\uparrow$ 11.53\%)\end{tabular}}   & \multicolumn{1}{c|}{\textbf{\begin{tabular}[c]{@{}c@{}}0.2839*\\ ($\uparrow$ 2.90\%)\end{tabular}}}     &   \textbf{\begin{tabular}[c]{@{}c@{}}0.2064*\\ ($\uparrow$ 3.82\%)\end{tabular}}   & \multicolumn{1}{c|}{\textbf{\begin{tabular}[c]{@{}c@{}}0.2909*\\ ($\uparrow$ 11.12\%)\end{tabular}}}     &  \textbf{\begin{tabular}[c]{@{}c@{}}0.2048*\\ ($\uparrow$ 11.67\%)\end{tabular}}    \\ \hline
\multicolumn{13}{|c|}{NeuMF} \\
\hline
BPR                     & \multicolumn{1}{c|}{0.0249}     &  0.0161    & \multicolumn{1}{c|}{0.1103}     &  0.0640    & \multicolumn{1}{c|}{0.1355}     & 0.0847     & \multicolumn{1}{c|}{0.0487}     &  0.0253    & \multicolumn{1}{c|}{0.1611}     & 0.1172     & \multicolumn{1}{c|}{0.1412}     &  0.1038    \\ \hline
WBPR                    & \multicolumn{1}{c|}{0.0272}     &  0.0174    & \multicolumn{1}{c|}{0.1105}     &  0.0662    & \multicolumn{1}{c|}{0.1435}     &   0.0860   & \multicolumn{1}{c|}{0.0496}     &  0.0267    & \multicolumn{1}{c|}{0.1629}     &   0.1191   & \multicolumn{1}{c|}{0.1431}     & 0.1051      \\ \hline
AOBPR                   & \multicolumn{1}{c|}{0.0263}     & 0.0166     & \multicolumn{1}{c|}{0.1108}     & 0.0643     & \multicolumn{1}{c|}{0.1359}     &   0.0852   & \multicolumn{1}{c|}{0.0515}     &  0.0271    & \multicolumn{1}{c|}{0.1677}     &  0.1328    & \multicolumn{1}{c|}{0.1597}     &  0.1162    \\ \hline
PRIS                    & \multicolumn{1}{c|}{0.0345}     &  0.0243   & \multicolumn{1}{c|}{0.1193}     &  0.0722    & \multicolumn{1}{c|}{0.1583}     &  0.0923    & \multicolumn{1}{c|}{0.0562}     &  0.0326    & \multicolumn{1}{c|}{0.1936}     &  0.1507    & \multicolumn{1}{c|}{0.2005}     &  0.1454    \\ \hline
TCE                     & \multicolumn{1}{c|}{0.0307}     & 0.0202     & \multicolumn{1}{c|}{0.1064}     & 0.0624     & \multicolumn{1}{c|}{0.1362}     &  0.0842    & \multicolumn{1}{c|}{0.0548}     & 0.0294     & \multicolumn{1}{c|}{0.1795}     &  0.1423    & \multicolumn{1}{c|}{0.1648}     &  0.1203    \\ \hline
RCE                     & \multicolumn{1}{c|}{0.0298}     & 0.0196     & \multicolumn{1}{c|}{0.1005}     &  0.0587    & \multicolumn{1}{c|}{0.1315}     &  0.0837    & \multicolumn{1}{c|}{0.0539}     & 0.0288     & \multicolumn{1}{c|}{0.1714}     &  0.1352    & \multicolumn{1}{c|}{0.1629}     &  0.1198    \\ \hline
TCE-BPR                 & \multicolumn{1}{c|}{0.0395}     &  0.0268    & \multicolumn{1}{c|}{0.1177}     &  0.0686    & \multicolumn{1}{c|}{0.1480}     &  0.0896    & \multicolumn{1}{c|}{0.0596}     &  0.0347    & \multicolumn{1}{c|}{0.1886}     &  0.1478    & \multicolumn{1}{c|}{0.1832}     &  0.1346    \\ \hline
RCE-BPR                 & \multicolumn{1}{c|}{0.0382}     &  0.0259    & \multicolumn{1}{c|}{0.1142}     & 0.0677     & \multicolumn{1}{c|}{0.1476}     &  0.0877    & \multicolumn{1}{c|}{0.0584}     &  0.0339    & \multicolumn{1}{c|}{0.1860}     &  0.1469    & \multicolumn{1}{c|}{0.1824}     &   0.1329   \\ \hline
TIL-UI                  & \multicolumn{1}{c|}{0.0459}     & 0.0307     & \multicolumn{1}{c|}{0.1185}     & 0.0701     & \multicolumn{1}{c|}{0.1689}     & 0.1090     & \multicolumn{1}{c|}{0.0707}     &  0.0379   & \multicolumn{1}{c|}{0.2229}     & 0.1870     & \multicolumn{1}{c|}{0.2692}     & 0.1884     \\ \hline
TIL-MI                  & \multicolumn{1}{c|}{\underline{0.0476}}     & \underline{0.0321}     & \multicolumn{1}{c|}{\underline{0.1333}}     &  \underline{0.0783}    & \multicolumn{1}{c|}{\underline{0.1732}}     &  \underline{0.1117}    & \multicolumn{1}{c|}{\underline{0.0729}}     & \underline{0.0387}     & \multicolumn{1}{c|}{\underline{0.2276}}     & \underline{0.1884}     & \multicolumn{1}{c|}{\underline{0.2739}}     &   \underline{0.1940}   \\ \hline
ITIPR                 & \multicolumn{1}{c|}{\textbf{\begin{tabular}[c]{@{}c@{}}0.0548*\\ ($\uparrow$ 15.13\%)\end{tabular}}}     &   \textbf{\begin{tabular}[c]{@{}c@{}}0.0364*\\ ($\uparrow$ 13.40\%)\end{tabular}}   & \multicolumn{1}{c|}{\textbf{\begin{tabular}[c]{@{}c@{}}0.1474*\\ ($\uparrow$ 10.58\%)\end{tabular}}}     & \textbf{\begin{tabular}[c]{@{}c@{}}0.0855*\\ ($\uparrow$ 9.20\%)\end{tabular}}     & \multicolumn{1}{c|}{\textbf{\begin{tabular}[c]{@{}c@{}}0.1972*\\ ($\uparrow$ 13.86\%)\end{tabular}}}     &   \textbf{\begin{tabular}[c]{@{}c@{}}0.1268*\\ ($\uparrow$ 13.52\%)\end{tabular}}   & \multicolumn{1}{c|}{\textbf{\begin{tabular}[c]{@{}c@{}}0.0840*\\ ($\uparrow$ 15.23\%)\end{tabular}}}     &   \textbf{\begin{tabular}[c]{@{}c@{}}0.0435*\\ ($\uparrow$ 12.40\%)\end{tabular}}   & \multicolumn{1}{c|}{\textbf{\begin{tabular}[c]{@{}c@{}}0.2463*\\ ($\uparrow$ 8.22\%)\end{tabular}}}     &   \textbf{\begin{tabular}[c]{@{}c@{}}0.1980*\\ ($\uparrow$ 5.10\%)\end{tabular}}   & \multicolumn{1}{c|}{\textbf{\begin{tabular}[c]{@{}c@{}}0.2898*\\ ($\uparrow$ 5.81\%)\end{tabular}}}     &  \textbf{\begin{tabular}[c]{@{}c@{}}0.2036*\\ ($\uparrow$ 4.95\%)\end{tabular}}    \\ \hline
\multicolumn{13}{|c|}{NGCF} \\
\hline
BPR                     & \multicolumn{1}{c|}{0.0519}     &  0.0283    & \multicolumn{1}{c|}{0.1258}     &  0.0729    & \multicolumn{1}{c|}{0.1578}     & 0.0997     & \multicolumn{1}{c|}{0.0757}     &   0.0402   & \multicolumn{1}{c|}{0.2575}     & 0.1930     & \multicolumn{1}{c|}{0.2175}     &   0.1624   \\ \hline
WBPR                    & \multicolumn{1}{c|}{0.0525}     &  0.0317    & \multicolumn{1}{c|}{0.1251}     &  0.0713    & \multicolumn{1}{c|}{0.1591}     &   0.1012   & \multicolumn{1}{c|}{0.0792}     & 0.0421    & \multicolumn{1}{c|}{0.2551}     & 0.1916     & \multicolumn{1}{c|}{0.2169}     &   0.1620   \\ \hline
AOBPR                   & \multicolumn{1}{c|}{0.0537}     & 0.0323     & \multicolumn{1}{c|}{0.1264}     & 0.0732     & \multicolumn{1}{c|}{0.1608}     &   0.1029   & \multicolumn{1}{c|}{0.0813}     &  0.0437    & \multicolumn{1}{c|}{0.2571}     &  0.1922    & \multicolumn{1}{c|}{0.2270}     &  0.1673    \\ \hline
PRIS                    & \multicolumn{1}{c|}{0.0589}     &  0.0381    & \multicolumn{1}{c|}{0.1275}     &  0.0737    & \multicolumn{1}{c|}{0.1682}     &  0.1054    & \multicolumn{1}{c|}{0.0843}     &  0.0459    & \multicolumn{1}{c|}{0.2583}     & 0.1935     & \multicolumn{1}{c|}{0.2435}     &  0.1754    \\ \hline
TCE                     & \multicolumn{1}{c|}{0.0506}     & 0.0280     & \multicolumn{1}{c|}{0.1243}     & 0.0698     & \multicolumn{1}{c|}{0.1622}     &  0.1035    & \multicolumn{1}{c|}{0.0746}     &   0.0394   & \multicolumn{1}{c|}{0.2556}     &  0.1924    & \multicolumn{1}{c|}{0.2397}     &  0.1738    \\ \hline
RCE                     & \multicolumn{1}{c|}{0.0517}     &  0.0282    & \multicolumn{1}{c|}{0.1239}     &  0.0683    & \multicolumn{1}{c|}{0.1617}     &  0.1032    & \multicolumn{1}{c|}{0.0765}     &  0.0407    & \multicolumn{1}{c|}{0.2543}     &  0.1912    & \multicolumn{1}{c|}{0.2418}     &   0.1749   \\ \hline
TCE-BPR                 & \multicolumn{1}{c|}{0.0548}     &  0.0337    & \multicolumn{1}{c|}{0.1262}     &  0.0730    & \multicolumn{1}{c|}{0.1632}     &  0.1045    & \multicolumn{1}{c|}{0.0789}     &  0.0418    & \multicolumn{1}{c|}{0.2598}     &   0.1943   & \multicolumn{1}{c|}{0.2451}     &   0.1768   \\ \hline
RCE-BPR                 & \multicolumn{1}{c|}{0.0562}     &  0.0354    & \multicolumn{1}{c|}{0.1254}     & 0.0718     & \multicolumn{1}{c|}{0.1624}     &  0.1037    & \multicolumn{1}{c|}{0.0804}     & 0.0428     & \multicolumn{1}{c|}{0.2587}     &  0.1936    & \multicolumn{1}{c|}{0.2465}     &  0.1775    \\ \hline
TIL-UI                  & \multicolumn{1}{c|}{0.0640}     &  0.0429    & \multicolumn{1}{c|}{0.1296}     & 0.0765     & \multicolumn{1}{c|}{0.1747}     & 0.1102     & \multicolumn{1}{c|}{0.0895}     & 0.0495     & \multicolumn{1}{c|}{0.2665}     &   0.1974   & \multicolumn{1}{c|}{0.2755}     &  0.1932    \\ \hline
TIL-MI                  & \multicolumn{1}{c|}{\underline{0.0649}}     &  \underline{0.0436}    & \multicolumn{1}{c|}{\underline{0.1351}}     &  \underline{0.0794}    & \multicolumn{1}{c|}{\underline{0.1784}}     &  \underline{0.1128}    & \multicolumn{1}{c|}{\underline{0.0910}}     &  \underline{0.0501}    & \multicolumn{1}{c|}{\underline{0.2694}}     &   \underline{0.1998}   & \multicolumn{1}{c|}{\underline{0.2796}}     &  \underline{0.1968}    \\ \hline
ITIPR                 & \multicolumn{1}{c|}{\textbf{\begin{tabular}[c]{@{}c@{}}0.0716*\\ ($\uparrow$ 10.32\%)\end{tabular}}}     &   \textbf{\begin{tabular}[c]{@{}c@{}}0.0493*\\ ($\uparrow$ 13.07\%)\end{tabular}}   & \multicolumn{1}{c|}{\textbf{\begin{tabular}[c]{@{}c@{}}0.1427*\\ ($\uparrow$ 5.63\%)\end{tabular}}}     & \textbf{\begin{tabular}[c]{@{}c@{}}0.0851*\\ ($\uparrow$ 7.18\%)\end{tabular}}     & \multicolumn{1}{c|}{\textbf{\begin{tabular}[c]{@{}c@{}}0.1936*\\ ($\uparrow$ 8.52\%)\end{tabular}}}     &   \textbf{\begin{tabular}[c]{@{}c@{}}0.1219*\\ ($\uparrow$ 8.07\%)\end{tabular}}   & \multicolumn{1}{c|}{\textbf{\begin{tabular}[c]{@{}c@{}}0.0974*\\ ($\uparrow$ 7.03\%)\end{tabular}}}     &   \textbf{\begin{tabular}[c]{@{}c@{}}0.0558*\\ ($\uparrow$ 11.38\%)\end{tabular}}   & \multicolumn{1}{c|}{\textbf{\begin{tabular}[c]{@{}c@{}}0.2873*\\ ($\uparrow$ 6.64\%)\end{tabular}}}     &   \textbf{\begin{tabular}[c]{@{}c@{}}0.2192*\\ ($\uparrow$ 9.71\%)\end{tabular}}   & \multicolumn{1}{c|}{\textbf{\begin{tabular}[c]{@{}c@{}}0.2948*\\ ($\uparrow$ 5.44\%)\end{tabular}}}     &  \textbf{\begin{tabular}[c]{@{}c@{}}0.2087*\\ ($\uparrow$ 6.05\%)\end{tabular}}    \\ \hline
\multicolumn{13}{|c|}{LightGCN} \\
\hline
BPR                     & \multicolumn{1}{c|}{0.0501}     &  0.0273    & \multicolumn{1}{c|}{0.1340}     &  0.0790    & \multicolumn{1}{c|}{0.1773}     & 0.1116     & \multicolumn{1}{c|}{0.0773}     &  0.0424    & \multicolumn{1}{c|}{0.3228}     & 0.2455     & \multicolumn{1}{c|}{0.2896}     &   0.2145   \\ \hline
WBPR                    & \multicolumn{1}{c|}{0.0533}     &  0.0319    & \multicolumn{1}{c|}{0.1352}     &  0.0796    & \multicolumn{1}{c|}{0.1778}     &   0.1118   & \multicolumn{1}{c|}{0.0798}     & 0.0435     & \multicolumn{1}{c|}{0.3221}     & 0.2465     & \multicolumn{1}{c|}{0.2875}     &  0.2114    \\ \hline
AOBPR                   & \multicolumn{1}{c|}{0.0558}     & 0.0342     & \multicolumn{1}{c|}{0.1344}     & 0.0795     & \multicolumn{1}{c|}{0.1775}     &   0.1121   & \multicolumn{1}{c|}{0.0833}     &   0.0441   & \multicolumn{1}{c|}{0.3210}     &  0.2451    & \multicolumn{1}{c|}{0.2889}     &  0.2132    \\ \hline
PRIS                    & \multicolumn{1}{c|}{0.0637}     &  0.0423    & \multicolumn{1}{c|}{0.1377}     &  0.0815    & \multicolumn{1}{c|}{\underline{0.1903}}     &  0.1130    & \multicolumn{1}{c|}{0.0852}     &  0.0465    & \multicolumn{1}{c|}{0.3259}     &  0.2483    & \multicolumn{1}{c|}{0.2934}     & 0.2156     \\ \hline
TCE                     & \multicolumn{1}{c|}{0.0586}     &  0.0379    & \multicolumn{1}{c|}{0.1294}     & 0.0759     & \multicolumn{1}{c|}{0.1782}     &  0.1124    & \multicolumn{1}{c|}{0.0746}     &  0.0409    & \multicolumn{1}{c|}{0.3234}     &  0.2457    & \multicolumn{1}{c|}{0.2893}     &  0.2141    \\ \hline
RCE                     & \multicolumn{1}{c|}{0.0575}     &  0.0368    & \multicolumn{1}{c|}{0.1286}     &  0.0721    & \multicolumn{1}{c|}{0.1772}     &  0.1112    & \multicolumn{1}{c|}{0.0758}     &   0.0413   & \multicolumn{1}{c|}{0.3226}     &  0.2453    & \multicolumn{1}{c|}{0.2880}     &  0.2128    \\ \hline
TCE-BPR                 & \multicolumn{1}{c|}{0.0613}     & 0.0405     & \multicolumn{1}{c|}{0.1352}     &  0.0812    & \multicolumn{1}{c|}{0.1799}     &  0.1121    & \multicolumn{1}{c|}{0.0839}     &  0.0443    & \multicolumn{1}{c|}{0.3247}     &  0.2480    & \multicolumn{1}{c|}{0.2932}     &  0.2153    \\ \hline
RCE-BPR                 & \multicolumn{1}{c|}{0.0601}     &  0.0395    & \multicolumn{1}{c|}{0.1345}     & 0.0798     & \multicolumn{1}{c|}{0.1782}     &  0.1115    & \multicolumn{1}{c|}{0.0814}     &  0.0439    & \multicolumn{1}{c|}{0.3237}     & 0.2459     & \multicolumn{1}{c|}{0.2917}     &  0.2147    \\ \hline
TIL-UI                  & \multicolumn{1}{c|}{0.0752}     & 0.0510     & \multicolumn{1}{c|}{0.1392}     & 0.0834     & \multicolumn{1}{c|}{0.1801}     & 0.1127     & \multicolumn{1}{c|}{0.0918}     & 0.0498     & \multicolumn{1}{c|}{0.3242}     & 0.2478     & \multicolumn{1}{c|}{0.2979}     &  0.2187    \\ \hline
TIL-MI                  & \multicolumn{1}{c|}{\underline{0.0762}}     &  \underline{0.0519}    & \multicolumn{1}{c|}{\underline{0.1421}}     &  \underline{0.0861}    & \multicolumn{1}{c|}{0.1861}     &  \underline{0.1162}    & \multicolumn{1}{c|}{\underline{0.0926}}     &  \underline{0.0507}    & \multicolumn{1}{c|}{\underline{0.3279}}     &  \underline{0.2491}    & \multicolumn{1}{c|}{\underline{0.2991}}     & \underline{0.2198}     \\ \hline
ITIPR                 & \multicolumn{1}{c|}{\textbf{\begin{tabular}[c]{@{}c@{}}0.0941*\\ ($\uparrow$ 23.49\%)\end{tabular}}}     &   \textbf{\begin{tabular}[c]{@{}c@{}}0.0676*\\ ($\uparrow$ 30.25\%)\end{tabular}}   & \multicolumn{1}{c|}{\textbf{\begin{tabular}[c]{@{}c@{}}0.1594*\\ ($\uparrow$ 12.17\%)\end{tabular}}}     & \textbf{\begin{tabular}[c]{@{}c@{}}0.0942*\\ ($\uparrow$ 9.41\%)\end{tabular}}     & \multicolumn{1}{c|}{\textbf{\begin{tabular}[c]{@{}c@{}}0.1998*\\ ($\uparrow$ 7.36\%)\end{tabular}}}     &   \textbf{\begin{tabular}[c]{@{}c@{}}0.1235*\\ ($\uparrow$ 6.28\%)\end{tabular}}   & \multicolumn{1}{c|}{\textbf{\begin{tabular}[c]{@{}c@{}}0.0994*\\ ($\uparrow$ 7.34\%)\end{tabular}}}     &   \textbf{\begin{tabular}[c]{@{}c@{}}0.0563*\\ ($\uparrow$ 11.05\%)\end{tabular}}   & \multicolumn{1}{c|}{\textbf{\begin{tabular}[c]{@{}c@{}}0.3440*\\ ($\uparrow$ 4.91\%)\end{tabular}}}     &   \textbf{\begin{tabular}[c]{@{}c@{}}0.2638*\\ ($\uparrow$ 5.90\%)\end{tabular}}   & \multicolumn{1}{c|}{\textbf{\begin{tabular}[c]{@{}c@{}}0.3128*\\ ($\uparrow$ 4.58\%)\end{tabular}}}     &  \textbf{\begin{tabular}[c]{@{}c@{}}0.2267*\\ ($\uparrow$ 3.14\%)\end{tabular}}    \\ \hline
\end{tabular}}
\label{table:overall performance}
\end{table*}

\subsection{Experiment Setup}
\subsubsection{\textbf{Datasets}}
We evaluate our method on six public datasets: Amazon-Books~\citep{he2016ups}, Amazon-CDs~\citep{he2016ups}, Yelp~\citep{asghar2016yelp}, Gowalla~\citep{cho2011friendship}, ML-20M~\citep{harper2015movielens}, and ML-Latest~\citep{harper2015movielens}. To accommodate the implicit feedback setting, for the datasets with explicit ratings, we preserve the ratings more than three (out of five) as positive feedback and treat all other ratings as missing entries.

\begin{table*}[]
\caption{The performance comparison between ITIPR(w/o TSWPL), ITIPR(w/o TSAR), and ITIPR with different base models on six datasets in terms of \textit{R@20(Recall@20)} and \textit{N@20(NDCG@20)}. The best results are marked in bold. $\downarrow$ denotes the performance degradation compared with the best method. $*$ indicates the improvements are statistically significant ($t$-test, $p$-value $\leq 0.01$).}
\vspace{-2mm}
\resizebox{0.97\textwidth}{!}{
\begin{tabular}{|c|cc|cc|cc|cc|cc|cc|}
\hline
\multirow{2}{*}{Methods} & \multicolumn{2}{c|}{Amazon-Books}       & \multicolumn{2}{c|}{Amazon-CDs}         & \multicolumn{2}{c|}{Gowalla}     & \multicolumn{2}{c|}{Yelp}        & \multicolumn{2}{c|}{ML-20M}   & \multicolumn{2}{c|}{ML-Latest}      \\ \cline{2-13}
                        & \multicolumn{1}{c|}{R@20} & N@20 & \multicolumn{1}{c|}{R@20} & N@20 & \multicolumn{1}{c|}{R@20} & N@20 & \multicolumn{1}{c|}{R@20} & N@20 & \multicolumn{1}{c|}{R@20} & N@20 & \multicolumn{1}{c|}{R@20} & N@20 \\ \hline
                        \multicolumn{13}{|c|}{MF} \\\hline
\multicolumn{1}{|c|}{\begin{tabular}[c]{@{}c@{}}ITIPR\\ (w/o TSWPL)\end{tabular}}                 & \multicolumn{1}{c|}{\begin{tabular}[c]{@{}c@{}}0.0623\\ ($\downarrow$ 7.70\%)\end{tabular}}   & \multicolumn{1}{c|}{\begin{tabular}[c]{@{}c@{}}0.0403\\ ($\downarrow$ 12.20\%)\end{tabular}}     & \multicolumn{1}{c|}{\begin{tabular}[c]{@{}c@{}}0.1129\\ ($\downarrow$ 11.93\%)\end{tabular}}     & \multicolumn{1}{c|}{\begin{tabular}[c]{@{}c@{}}0.0654\\ ($\downarrow$ 15.55\%)\end{tabular}}     & \multicolumn{1}{c|}{\begin{tabular}[c]{@{}c@{}}0.1625\\ ($\downarrow$ 9.01\%)\end{tabular}}     & \multicolumn{1}{c|}{\begin{tabular}[c]{@{}c@{}}0.0995\\ ($\downarrow$ 7.10\%)\end{tabular}}     & \multicolumn{1}{c|}{\begin{tabular}[c]{@{}c@{}}0.0926\\ ($\downarrow$ 10.27\%)\end{tabular}}     & \multicolumn{1}{c|}{\begin{tabular}[c]{@{}c@{}}0.0503\\ ($\downarrow$ 10.34\%)\end{tabular}}    & \multicolumn{1}{c|}{\begin{tabular}[c]{@{}c@{}}0.2681\\ ($\downarrow$ 5.57\%)\end{tabular}}     &  \multicolumn{1}{c|}{\begin{tabular}[c]{@{}c@{}}0.1952\\ ($\downarrow$ 5.43\%)\end{tabular}}    & \multicolumn{1}{c|}{\begin{tabular}[c]{@{}c@{}}0.2654\\ ($\downarrow$ 8.77\%)\end{tabular}}     & \multicolumn{1}{c|}{\begin{tabular}[c]{@{}c@{}}0.1825\\ ($\downarrow$ 10.89\%)\end{tabular}}     \\ \hline
\multicolumn{1}{|c|}{\begin{tabular}[c]{@{}c@{}}ITIPR\\ (w/o TSAR)\end{tabular}}                 & \multicolumn{1}{c|}{\begin{tabular}[c]{@{}c@{}}0.0586\\ ($\downarrow$ 14.51\%)\end{tabular}}     & \multicolumn{1}{c|}{\begin{tabular}[c]{@{}c@{}}0.0396\\ ($\downarrow$ 13.73\%)\end{tabular}}     & \multicolumn{1}{c|}{\begin{tabular}[c]{@{}c@{}}0.1152\\ ($\downarrow$ 10.14\%)\end{tabular}}     &  \multicolumn{1}{c|}{\begin{tabular}[c]{@{}c@{}}0.0678\\ ($\downarrow$ 9.12\%)\end{tabular}}   & \multicolumn{1}{c|}{\begin{tabular}[c]{@{}c@{}}0.1583\\ ($\downarrow$ 11.37\%)\end{tabular}}     &  \multicolumn{1}{c|}{\begin{tabular}[c]{@{}c@{}}0.0967\\ ($\downarrow$ 9.71\%)\end{tabular}}    & \multicolumn{1}{c|}{\begin{tabular}[c]{@{}c@{}}0.0873\\ ($\downarrow$ 15.41\%)\end{tabular}}     &  \multicolumn{1}{c|}{\begin{tabular}[c]{@{}c@{}}0.0482\\ ($\downarrow$ 14.08\%)\end{tabular}}    & \multicolumn{1}{c|}{\begin{tabular}[c]{@{}c@{}}0.2635\\ ($\downarrow$ 7.19\%)\end{tabular}}     &   \multicolumn{1}{c|}{\begin{tabular}[c]{@{}c@{}}0.1936\\ ($\downarrow$ 6.20\%)\end{tabular}}   & \multicolumn{1}{c|}{\begin{tabular}[c]{@{}c@{}}0.2627\\ ($\downarrow$ 9.69\%)\end{tabular}}     & \multicolumn{1}{c|}{\begin{tabular}[c]{@{}c@{}}0.1807\\ ($\downarrow$ 11.77\%)\end{tabular}}     \\ \hline
ITIPR                 & \multicolumn{1}{c|}{\textbf{0.0671*}}     &   \textbf{0.0459*}   & \multicolumn{1}{c|}{\textbf{0.1282*}}     & \textbf{0.0746*}     & \multicolumn{1}{c|}{\textbf{0.1786*}}     &   \textbf{0.1071*}  & \multicolumn{1}{c|}{\textbf{0.1032*}}    &   \textbf{0.0561*}   & \multicolumn{1}{c|}{\textbf{0.2839*}}    &   \textbf{0.2064*}   & \multicolumn{1}{c|}{\textbf{0.2909*}}     &  \textbf{0.2048*}     \\ \hline
\multicolumn{13}{|c|}{NeuMF} \\
\hline
\multicolumn{1}{|c|}{\begin{tabular}[c]{@{}c@{}}ITIPR\\ (w/o TSWPL)\end{tabular}}                   & \multicolumn{1}{c|}{\begin{tabular}[c]{@{}c@{}}0.0472\\ ($\downarrow$ 13.87\%)\end{tabular}}     & \multicolumn{1}{c|}{\begin{tabular}[c]{@{}c@{}}0.0322\\ ($\downarrow$ 11.54\%)\end{tabular}}     & \multicolumn{1}{c|}{\begin{tabular}[c]{@{}c@{}}0.1348\\ ($\downarrow$ 8.55\%)\end{tabular}}     & \multicolumn{1}{c|}{\begin{tabular}[c]{@{}c@{}}0.0746\\ ($\downarrow$ 12.75\%)\end{tabular}}     & \multicolumn{1}{c|}{\begin{tabular}[c]{@{}c@{}}0.1778\\ ($\downarrow$ 9.84\%)\end{tabular}}     & \multicolumn{1}{c|}{\begin{tabular}[c]{@{}c@{}}0.1123\\ ($\downarrow$ 11.44\%)\end{tabular}}     & \multicolumn{1}{c|}{\begin{tabular}[c]{@{}c@{}}0.0728\\ ($\downarrow$ 13.33\%)\end{tabular}}     &  \multicolumn{1}{c|}{\begin{tabular}[c]{@{}c@{}}0.0394\\ ($\downarrow$ 9.43\%)\end{tabular}}   & \multicolumn{1}{c|}{\begin{tabular}[c]{@{}c@{}}0.2281\\ ($\downarrow$ 7.39\%)\end{tabular}}     & \multicolumn{1}{c|}{\begin{tabular}[c]{@{}c@{}}0.1885\\ ($\downarrow$ 4.80\%)\end{tabular}}     & \multicolumn{1}{c|}{\begin{tabular}[c]{@{}c@{}}0.2720\\ ($\downarrow$ 6.14\%)\end{tabular}}     & \multicolumn{1}{c|}{\begin{tabular}[c]{@{}c@{}}0.1902\\ ($\downarrow$6.58\%)\end{tabular}}     \\ \hline
\multicolumn{1}{|c|}{\begin{tabular}[c]{@{}c@{}}ITIPR\\ (w/o TSAR)\end{tabular}}                  & \multicolumn{1}{c|}{\begin{tabular}[c]{@{}c@{}}0.0454\\ ($\downarrow$ 17.15\%)\end{tabular}}     & \multicolumn{1}{c|}{\begin{tabular}[c]{@{}c@{}}0.0315\\ ($\downarrow$ 13.46\%)\end{tabular}}     & \multicolumn{1}{c|}{\begin{tabular}[c]{@{}c@{}}0.1311\\ ($\downarrow$ 11.06\%)\end{tabular}}     &  \multicolumn{1}{c|}{\begin{tabular}[c]{@{}c@{}}0.0729\\ ($\downarrow$ 14.74\%)\end{tabular}}    & \multicolumn{1}{c|}{\begin{tabular}[c]{@{}c@{}}0.1763\\ ($\downarrow$ 10.60\%)\end{tabular}}     &  \multicolumn{1}{c|}{\begin{tabular}[c]{@{}c@{}}0.1097\\ ($\downarrow$ 13.49\%)\end{tabular}}    & \multicolumn{1}{c|}{\begin{tabular}[c]{@{}c@{}}0.0745\\ ($\downarrow$ 11.31\%)\end{tabular}}     & \multicolumn{1}{c|}{\begin{tabular}[c]{@{}c@{}}0.0401\\ ($\downarrow$ 7.82\%)\end{tabular}}     & \multicolumn{1}{c|}{\begin{tabular}[c]{@{}c@{}}0.2234\\ ($\downarrow$ 9.30\%)\end{tabular}}     & \multicolumn{1}{c|}{\begin{tabular}[c]{@{}c@{}}0.1845\\ ($\downarrow$ 6.82\%)\end{tabular}}     & \multicolumn{1}{c|}{\begin{tabular}[c]{@{}c@{}}0.2692\\ ($\downarrow$ 7.11\%)\end{tabular}}     &   \multicolumn{1}{c|}{\begin{tabular}[c]{@{}c@{}}0.1867\\ ($\downarrow$ 8.30\%)\end{tabular}}   \\ \hline
ITIPR                 & \multicolumn{1}{c|}{\textbf{0.0548*}}     &   \textbf{0.0364*}   & \multicolumn{1}{c|}{\textbf{0.1474*}}     & \textbf{0.0855*}     & \multicolumn{1}{c|}{\textbf{0.1972*}}     &   \textbf{0.1268*}   & \multicolumn{1}{c|}{\textbf{0.0840*}}     &   \textbf{0.0435*}   & \multicolumn{1}{c|}{\textbf{0.2463*}}     &   \textbf{0.1980*}   & \multicolumn{1}{c|}{\textbf{0.2898*}}     &  \textbf{0.2036*}    \\ \hline
\multicolumn{13}{|c|}{NGCF} \\
\hline
\multicolumn{1}{|c|}{\begin{tabular}[c]{@{}c@{}}ITIPR\\ (w/o TSWPL)\end{tabular}}                   & \multicolumn{1}{c|}{\begin{tabular}[c]{@{}c@{}}0.0656\\ ($\downarrow$ 8.38\%)\end{tabular}}     & \multicolumn{1}{c|}{\begin{tabular}[c]{@{}c@{}}0.0445\\ ($\downarrow$ 9.74\%)\end{tabular}}     & \multicolumn{1}{c|}{\begin{tabular}[c]{@{}c@{}}0.1318\\ ($\downarrow$ 7.64\%)\end{tabular}}     & \multicolumn{1}{c|}{\begin{tabular}[c]{@{}c@{}}0.0783\\ ($\downarrow$ 7.99\%)\end{tabular}}     & \multicolumn{1}{c|}{\begin{tabular}[c]{@{}c@{}}0.1772\\ ($\downarrow$ 8.47\%)\end{tabular}}     & \multicolumn{1}{c|}{\begin{tabular}[c]{@{}c@{}}0.1136\\ ($\downarrow$ 6.81\%)\end{tabular}}     & \multicolumn{1}{c|}{\begin{tabular}[c]{@{}c@{}}0.0872\\ ($\downarrow$ 10.47\%)\end{tabular}}     &  \multicolumn{1}{c|}{\begin{tabular}[c]{@{}c@{}}0.0481\\ ($\downarrow$ 13.80\%)\end{tabular}}   & \multicolumn{1}{c|}{\begin{tabular}[c]{@{}c@{}}0.2641\\ ($\downarrow$ 8.08\%)\end{tabular}}     & \multicolumn{1}{c|}{\begin{tabular}[c]{@{}c@{}}0.1943\\ ($\downarrow$ 11.36\%)\end{tabular}}     & \multicolumn{1}{c|}{\begin{tabular}[c]{@{}c@{}}0.2733\\ ($\downarrow$ 7.29\%)\end{tabular}}     & \multicolumn{1}{c|}{\begin{tabular}[c]{@{}c@{}}0.1921\\ ($\downarrow$ 7.95\%)\end{tabular}}     \\ \hline
\multicolumn{1}{|c|}{\begin{tabular}[c]{@{}c@{}}ITIPR\\ (w/o TSAR)\end{tabular}}                  & \multicolumn{1}{c|}{\begin{tabular}[c]{@{}c@{}}0.0638\\ ($\downarrow$ 10.89\%)\end{tabular}}     & \multicolumn{1}{c|}{\begin{tabular}[c]{@{}c@{}}0.0427\\ ($\downarrow$ 13.39\%)\end{tabular}}     & \multicolumn{1}{c|}{\begin{tabular}[c]{@{}c@{}}0.1295\\ ($\downarrow$ 9.25\%)\end{tabular}}     &  \multicolumn{1}{c|}{\begin{tabular}[c]{@{}c@{}}0.0756\\ ($\downarrow$ 11.16\%)\end{tabular}}    & \multicolumn{1}{c|}{\begin{tabular}[c]{@{}c@{}}0.1754\\ ($\downarrow$ 9.40\%)\end{tabular}}     &  \multicolumn{1}{c|}{\begin{tabular}[c]{@{}c@{}}0.1117\\ ($\downarrow$ 8.37\%)\end{tabular}}    & \multicolumn{1}{c|}{\begin{tabular}[c]{@{}c@{}}0.0894\\ ($\downarrow$ 8.21\%)\end{tabular}}     & \multicolumn{1}{c|}{\begin{tabular}[c]{@{}c@{}}0.0503\\ ($\downarrow$ 9.87\%)\end{tabular}}     & \multicolumn{1}{c|}{\begin{tabular}[c]{@{}c@{}}0.2608\\ ($\downarrow$ 9.22\%)\end{tabular}}     & \multicolumn{1}{c|}{\begin{tabular}[c]{@{}c@{}}0.1917\\ ($\downarrow$ 12.55\%)\end{tabular}}     & \multicolumn{1}{c|}{\begin{tabular}[c]{@{}c@{}}0.2695\\ ($\downarrow$ 8.58\%)\end{tabular}}     &   \multicolumn{1}{c|}{\begin{tabular}[c]{@{}c@{}}0.1883\\ ($\downarrow$ 9.77\%)\end{tabular}}   \\ \hline
ITIPR                 & \multicolumn{1}{c|}{\textbf{0.0716*}}     &   \textbf{0.0493*}   & \multicolumn{1}{c|}{\textbf{0.1427*}}     & \textbf{0.0851*}     & \multicolumn{1}{c|}{\textbf{0.1936*}}     &   \textbf{0.1219*}   & \multicolumn{1}{c|}{\textbf{0.0974*}}     &   \textbf{0.0558*}   & \multicolumn{1}{c|}{\textbf{0.2873*}}     &   \textbf{0.2192*}   & \multicolumn{1}{c|}{\textbf{0.2948*}}     &  \textbf{0.2087*}    \\ \hline
\multicolumn{13}{|c|}{LightGCN} \\
\hline
\multicolumn{1}{|c|}{\begin{tabular}[c]{@{}c@{}}ITIPR\\ (w/o TSWPL)\end{tabular}}                   & \multicolumn{1}{c|}{\begin{tabular}[c]{@{}c@{}}0.0823\\ ($\downarrow$ 12.54\%)\end{tabular}}     & \multicolumn{1}{c|}{\begin{tabular}[c]{@{}c@{}}0.0585\\ ($\downarrow$ 13.46\%)\end{tabular}}     & \multicolumn{1}{c|}{\begin{tabular}[c]{@{}c@{}}0.1436\\ ($\downarrow$ 9.91\%)\end{tabular}}     & \multicolumn{1}{c|}{\begin{tabular}[c]{@{}c@{}}0.0866\\ ($\downarrow$ 8.07\%)\end{tabular}}     & \multicolumn{1}{c|}{\begin{tabular}[c]{@{}c@{}}0.1852\\ ($\downarrow$ 7.31\%)\end{tabular}}     & \multicolumn{1}{c|}{\begin{tabular}[c]{@{}c@{}}0.1169\\ ($\downarrow$ 5.34\%)\end{tabular}}     & \multicolumn{1}{c|}{\begin{tabular}[c]{@{}c@{}}0.0921\\ ($\downarrow$ 7.34\%)\end{tabular}}     &  \multicolumn{1}{c|}{\begin{tabular}[c]{@{}c@{}}0.0494\\ ($\downarrow$ 12.26\%)\end{tabular}}   & \multicolumn{1}{c|}{\begin{tabular}[c]{@{}c@{}}0.3281\\ ($\downarrow$ 4.62\%)\end{tabular}}     & \multicolumn{1}{c|}{\begin{tabular}[c]{@{}c@{}}0.2462\\ ($\downarrow$ 6.67\%)\end{tabular}}     & \multicolumn{1}{c|}{\begin{tabular}[c]{@{}c@{}}0.2987\\ ($\downarrow$ 4.51\%)\end{tabular}}     & \multicolumn{1}{c|}{\begin{tabular}[c]{@{}c@{}}0.2150\\ ($\downarrow$ 5.16\%)\end{tabular}}     \\ \hline
\multicolumn{1}{|c|}{\begin{tabular}[c]{@{}c@{}}ITIPR\\ (w/o TSAR)\end{tabular}}                  & \multicolumn{1}{c|}{\begin{tabular}[c]{@{}c@{}}0.0798\\ ($\downarrow$ 15.20\%)\end{tabular}}     & \multicolumn{1}{c|}{\begin{tabular}[c]{@{}c@{}}0.0558\\ ($\downarrow$ 17.46\%)\end{tabular}}     & \multicolumn{1}{c|}{\begin{tabular}[c]{@{}c@{}}0.1418\\ ($\downarrow$ 11.04\%)\end{tabular}}     &  \multicolumn{1}{c|}{\begin{tabular}[c]{@{}c@{}}0.0847\\ ($\downarrow$ 10.08\%)\end{tabular}}    & \multicolumn{1}{c|}{\begin{tabular}[c]{@{}c@{}}0.1834\\ ($\downarrow$ 8.21\%)\end{tabular}}     &  \multicolumn{1}{c|}{\begin{tabular}[c]{@{}c@{}}0.1134\\ ($\downarrow$ 8.18\%)\end{tabular}}    & \multicolumn{1}{c|}{\begin{tabular}[c]{@{}c@{}}0.0947\\ ($\downarrow$ 4.73\%)\end{tabular}}     & \multicolumn{1}{c|}{\begin{tabular}[c]{@{}c@{}}0.0510\\ ($\downarrow$ 9.41\%)\end{tabular}}     & \multicolumn{1}{c|}{\begin{tabular}[c]{@{}c@{}}0.3246\\ ($\downarrow$ 5.64\%)\end{tabular}}     & \multicolumn{1}{c|}{\begin{tabular}[c]{@{}c@{}}0.2439\\ ($\downarrow$ 7.54\%)\end{tabular}}     & \multicolumn{1}{c|}{\begin{tabular}[c]{@{}c@{}}0.2963\\ ($\downarrow$ 5.27\%)\end{tabular}}     &   \multicolumn{1}{c|}{\begin{tabular}[c]{@{}c@{}}0.2091\\ ($\downarrow$ 7.76\%)\end{tabular}}   \\ \hline
ITIPR                 & \multicolumn{1}{c|}{\textbf{0.0941*}}     &   \textbf{0.0676*}   & \multicolumn{1}{c|}{\textbf{0.1594*}}     & \textbf{0.0942*}     & \multicolumn{1}{c|}{\textbf{0.1998*}}     &   \textbf{0.1235*}   & \multicolumn{1}{c|}{\textbf{0.0994*}}     &   \textbf{0.0563*}   & \multicolumn{1}{c|}{\textbf{0.3440*}}     &   \textbf{0.2638*}   & \multicolumn{1}{c|}{\textbf{0.3128*}}     &  \textbf{0.2267*}    \\ \hline
\end{tabular}}
\label{table:ablation study}
\end{table*}
\vspace{-1mm}

\subsubsection{\textbf{Baselines}}
Generally, previous works on utilizing implicit feedback for personalized ranking can be divided into four categories. Note that we do not compare with works on \textit{using auxiliary information} considering this kind of information is often inaccessible. In experiments, we use representative works in left three categories as our baselines:
\vspace{1mm}

\textit{Negative item sampling:}
\begin{itemize}[leftmargin=*]
    \item \textbf{WBPR~\citep{gantner2012personalized}:} This method assigns more sampling probabilities to the negative items with higher global popularity.
    \item \textbf{AOBPR~\citep{rendle2014improving}:} It adopts a context-specific adaptive sampler to oversample popular negative items.
    \item \textbf{PRIS~\citep{lian2020personalized}:} It utilizes the importance sampling to obtain negative samples according to their informativeness. 
\end{itemize}
\textit{Positive pair re-weighting}:
\begin{itemize}[leftmargin=*]
    \item \textbf{TCE} and \textbf{RCE~\citep{wang2021denoising}:} They adaptively prune noisy positive interactions with large loss values in the training.   
    \item \textbf{TCE-BPR} and \textbf{RCE-BPR:}  The modified versions of the above two methods in which the point-wise loss is replaced with a pair-wise ranking loss following BPR's scheme.
\end{itemize}

\textit{Triplet importance modeling}:
\begin{itemize}[leftmargin=*]
    \item \textbf{TIL-UI} and \textbf{TIL-MI~\citep{wu2022adapting}:} They design a neural network-based weight generation module to learn triplet importance.  
\end{itemize}
Besides, we also utilize vanilla \textbf{BPR} as our initial baseline, which samples the negative samples uniformly and assigns identical importance to each triplet in the training stage. 
As for the recommendation model backbone, we employ MF~\citep{koren2009matrix} and NeuMF~\citep{he2017neural} as MF-based models, alongside NGCF~\citep{wang2019neural} and LightGCN~\citep{he2020lightgcn} as GNN-based models.
\vspace{-1mm}
\subsubsection{\textbf{Evaluation Metrics}}
We evaluate baselines and our method in terms of \textit{Recall@k} and \textit{NDCG@k}. For each user, the recommendation model will recommend an ordered list of items to her. \textit{Recall@k} (abbreviated as \textit{R@k}) indicates the percentage of her rated items that appear in the top-$k$ items of the recommended list. The \textit{NDCG@k} (abbreviated as \textit{N@k}) is the normalized discounted cumulative gain at a ranking position $k$ to measure the ranking quality. Similar to the previous related paper~\citep{wu2022adapting}, we set the $k$ as 20.


\subsubsection{\textbf{Implementation Details}}
We split user-item interaction records in each dataset randomly into the training set, validation set, and test set with the ratio of 8:1:1. In the implementation, we run each experiment five times with different seeds and report the average result as the final performance. In each individual experiment, regardless of the method chosen, the recommendation model is trained for a total of 1000 epochs. The early stopping scheme has also been adopted to prevent overfitting. We implement our method with PyTorch and run it on an NVIDIA RTX A6000 GPU with 48GB memory. 
We provide the code in the following repository to facilitate reproduction: \textcolor{blue}{\url{https://github.com/BokwaiHo/ITIPR.git}}.
\vspace{-2mm}

\subsection{Results and Analysis}
\subsubsection{\textbf{Overall Performance (RQ1)}}
We provide the experiment results of all baselines and our ITIPR with different base recommendation models on six public datasets in Tabel~\ref{table:overall performance}. First, it can be noticed that \textit{triplet importance modeling} methods: TIL-UI, TIL-MI, ITIPR achieve significantly better results than other types of methods, which validates the rationality of considering the personalized ranking problem from the triplet-level importance modeling perspective. Second, from the table, we can observe obvious performance improvement of our ITIPR over previous \textit{negative item sampling}, \textit{positive pair re-weighting}, \textit{using auxiliary information}, and \textit{triplet importance modeling} baselines in all public datasets on both \textit{Recall@20} and \textit{NDCG@20} metrics. In detail, ITIPR outperforms the best-performing baseline by $9.16\%$ and $9.62\%$ averagely on \textit{Recall@20} and \textit{NDCG@20} metrics, respectively, under all settings. This demonstrates that our triplet Shapley-based interpretable triplet importance framework for personalized ranking can effectively enhance the recommendation performance. 
\vspace{-1mm}

\subsubsection{\textbf{Applicability Analysis (RQ2)}}
The results presented in Table~\ref{table:overall performance} demonstrate the consistent improvement in recommendation performance achieved by our method compared with the baselines across various base recommendation models. Taking the \textit{Recall@20} metric as an example, in fact, our ITIPR achieves an average $7.94\%$ and $11.47\%$ improvement on MF and NeuMF base models, respectively. Meanwhile, for GNN-based models, the average gain of $7.26\%$ and $9.98\%$ can also be observed on NGCF and LightGCN, respectively. Similar improvement trends can also be noticed on another metric \textit{NDCG@20}. Therefore, no matter which mainstream base recommendation model we employ, ITIPR can effectively bring positive performance contributions.
\vspace{-1mm}

\subsubsection{\textbf{Ablation Study (RQ3)}}
To conduct the ablation study, we remove \textit{triplet Shapley-weighted pairwise learning} and \textit{triplet Shapley-aware resampling} from our overall ITIPR method respectively to obtain ITIPR(w/o TSWPR) and ITIPR(w/o TSAR). We compare their performance with ITIPR on all six datasets in Table~\ref{table:ablation study}. Taking an average over all base models and datasets, the performance of ITIPR(w/o TSWPR) is $8.71\%$ and $9.44\%$ lower than the complete ITIPR on \textit{Recall@20} and \textit{NDCG@20} metrics, respectively. The similar performance declines of $10.02\%$ and $10.62\%$ can also be noticed for ITIPR(w/o TSAR). Such results demonstrate that both \textit{triplet Shapley-weighted pairwise learning} and \textit{triplet Shapley-aware resampling} components contribute to the final performance effectively.
\vspace{-4mm}

\subsubsection{\textbf{Case Study (RQ4)}}
The main motivation of ITIPR is to provide an interpretable triplet importance measurement and develop performance enhancement methods based on this. To validate that the importance score assigned to each triplet is reasonable and intuitive, we visualize the triplet Shapley values for different samples from Amazon-CDs dataset in Table~\ref{table:case study}. According to the definition of triplet Shapley, for a triplet with a higher importance score, we use that it brings more contribution to the recommendation model performance as the explanation. From the table, we can find the triplets $1, 4, 7, 10$ exhibit higher triplet Shapley scores. This is consistent with the facts that the negative items in these triplets are more popular among the whole item set~\citep{lian2020personalized} or more different from the positive items~\citep{wu2022adapting} which can help the model better discern the user preference according to empirical experience and human expertise. Meanwhile, for the triplets with lower or even negative triplet Shapley values, their negative items are similar to corresponding positive ones, and some of them are from the same singer. Thus, we can conclude that our triplet Shapley-based explanations to the assigned importance scores are reasonable and intuitive.


\begin{table}[]
\caption{Triplet examples and corresponding normalized triplet Shapley values obtained with ITIPR on Amazon-CDs.}
\vspace{-1mm}
\resizebox{0.48\textwidth}{!}{
\begin{tabular}{|c|c|c|c|c|c|}
\hline
\multicolumn{1}{|c|}{\textbf{ID}} & \multicolumn{1}{|c|}{\textbf{Triplet}} & \multicolumn{1}{|c|}{\begin{tabular}[c]{@{}c@{}}\textbf{Triplet}\\ \textbf{Shapley}\end{tabular}} & \multicolumn{1}{|c|}{\textbf{ID}} & \multicolumn{1}{c|}{\textbf{Triplet}} & \multicolumn{1}{|c|}{\begin{tabular}[c]{@{}c@{}}\textbf{Triplet}\\ \textbf{Shapley}\end{tabular}} \\ \hline
1 & \raisebox{-0.7mm}{\includegraphics[width=1.7cm]{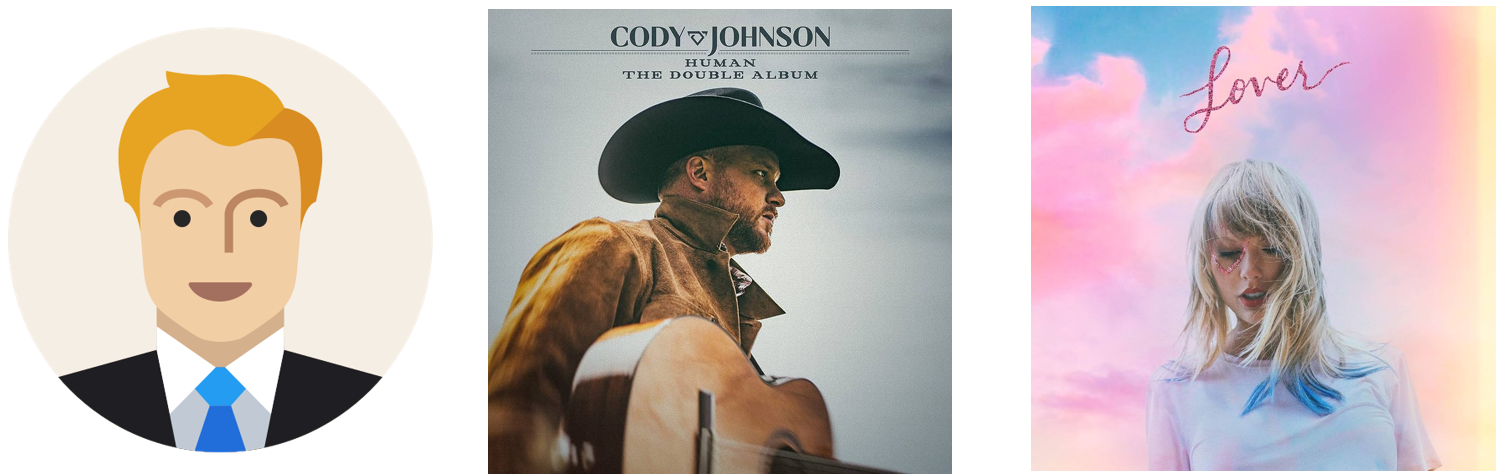}}    &      1.2865                                &   7 &     \raisebox{-0.7mm}{\includegraphics[width=1.7cm]{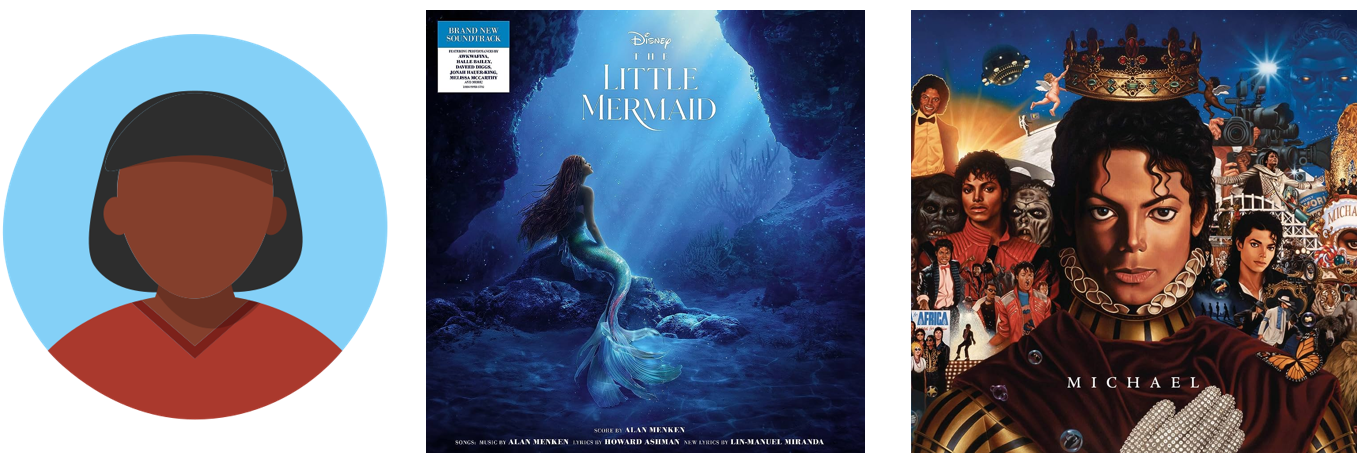}}                      &     0.9763            \\ \hline
2 & \raisebox{-0.7mm}{\includegraphics[width=1.7cm]{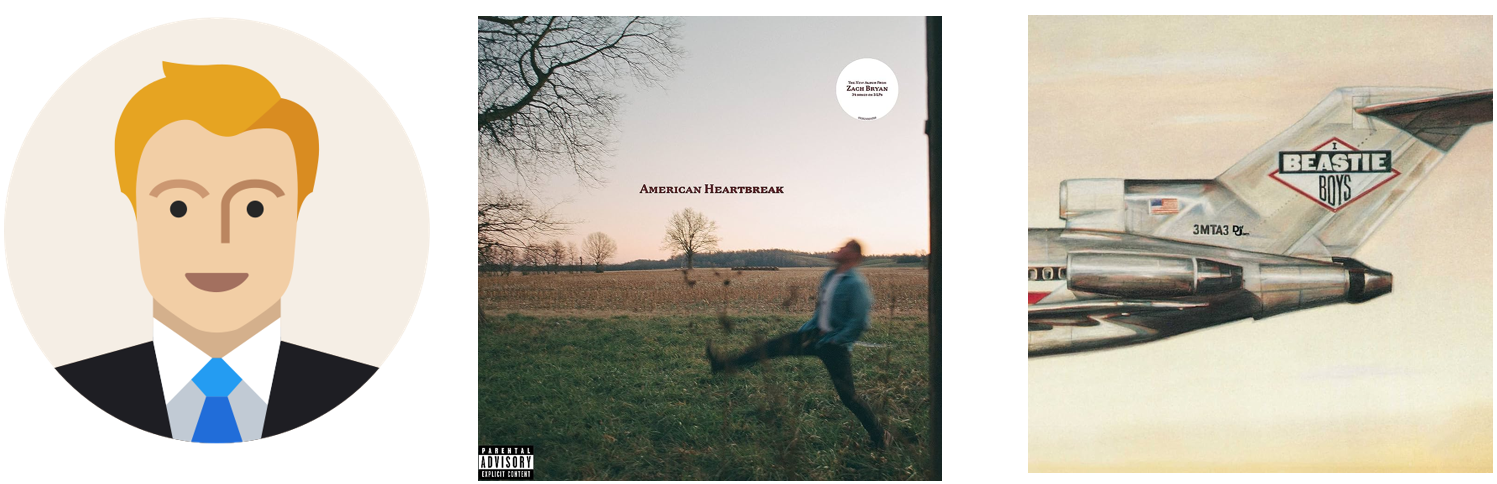}}                              &                       0.3470         & 8      &   \raisebox{-0.7mm}{\includegraphics[width=1.7cm]{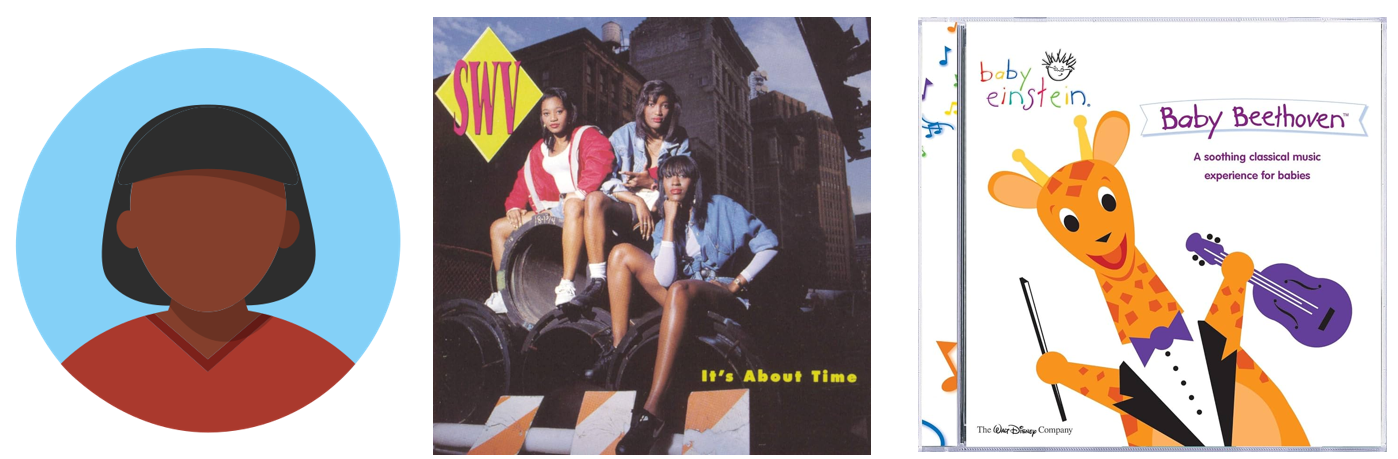}}                           &   0.6831              \\ \hline
3 & \raisebox{-0.7mm}{\includegraphics[width=1.7cm]{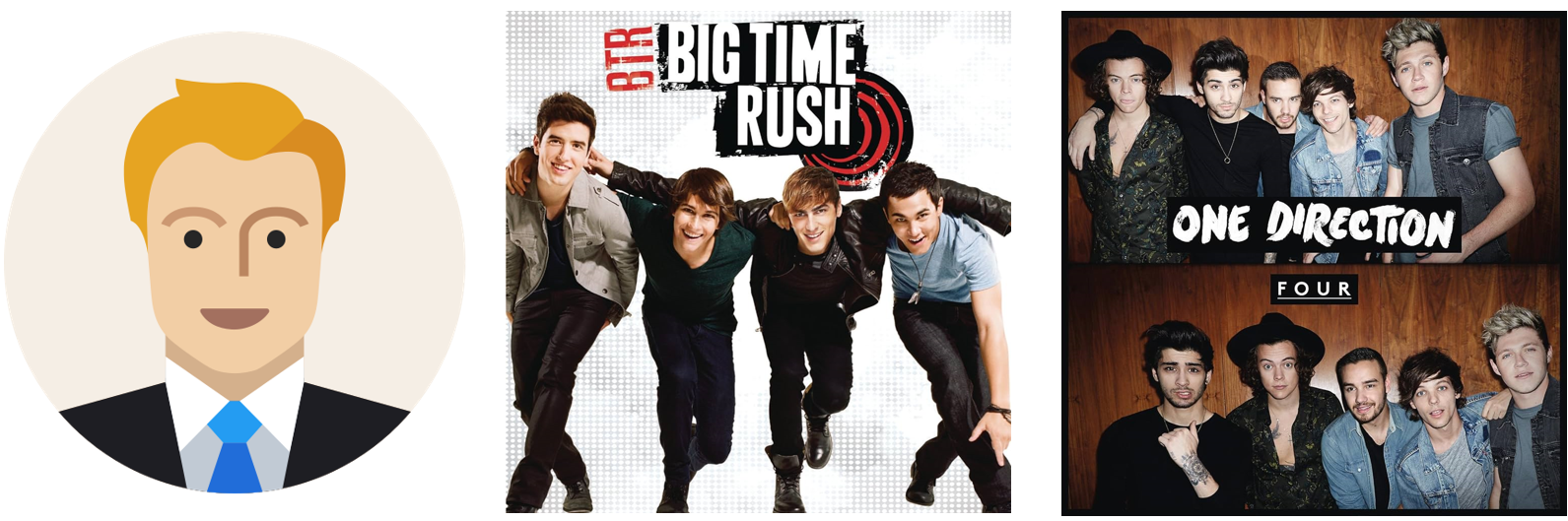}}                              &              0.0019          & 9              &  \raisebox{-0.7mm}{\includegraphics[width=1.7cm]{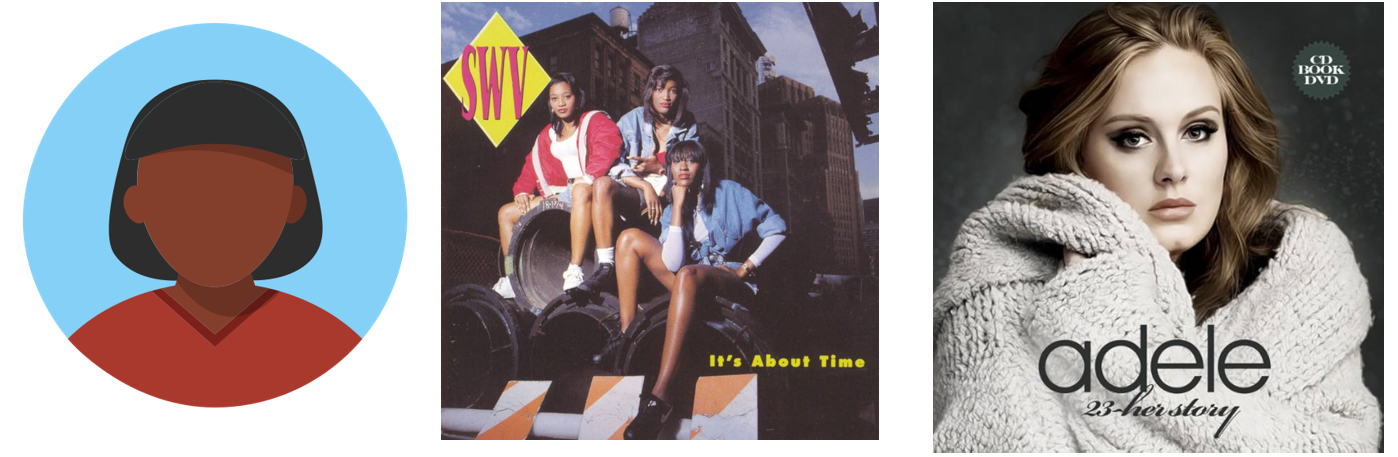}}                            &     0.0842            \\ \hline
4 & \raisebox{-0.7mm}{\includegraphics[width=1.7cm]{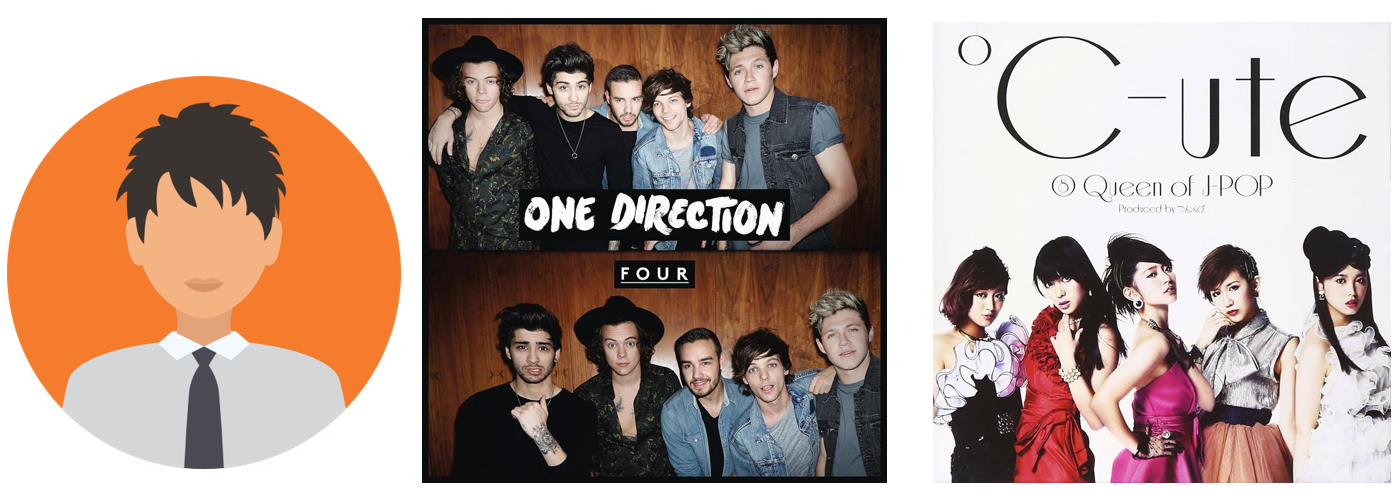}}                              &              1.4084          & 10              &   \raisebox{-0.7mm}{\includegraphics[width=1.7cm]{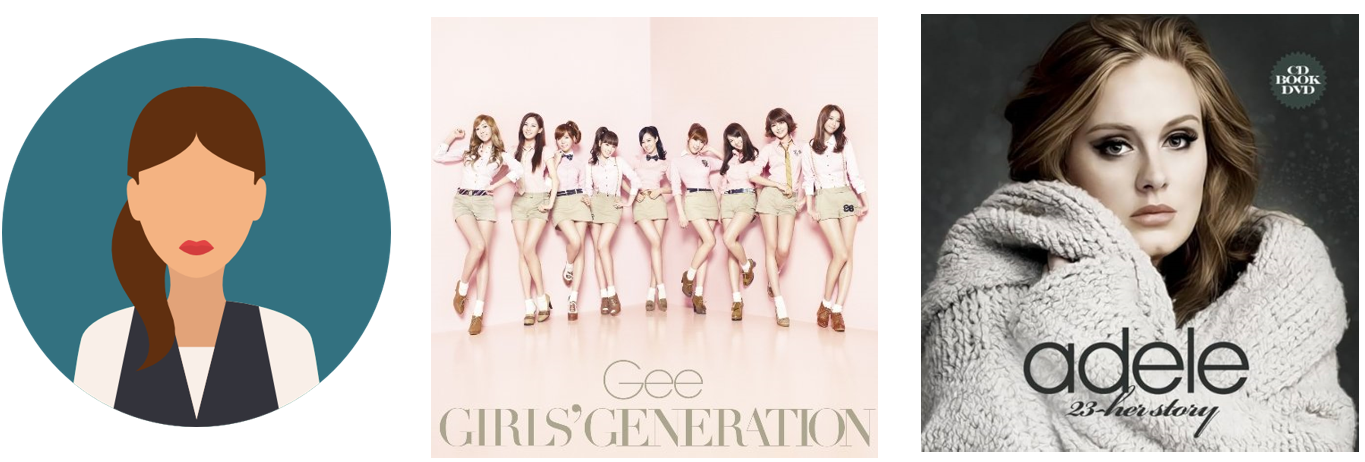}}                           &   0.8567              \\ \hline
5 & \raisebox{-0.7mm}{\includegraphics[width=1.7cm]{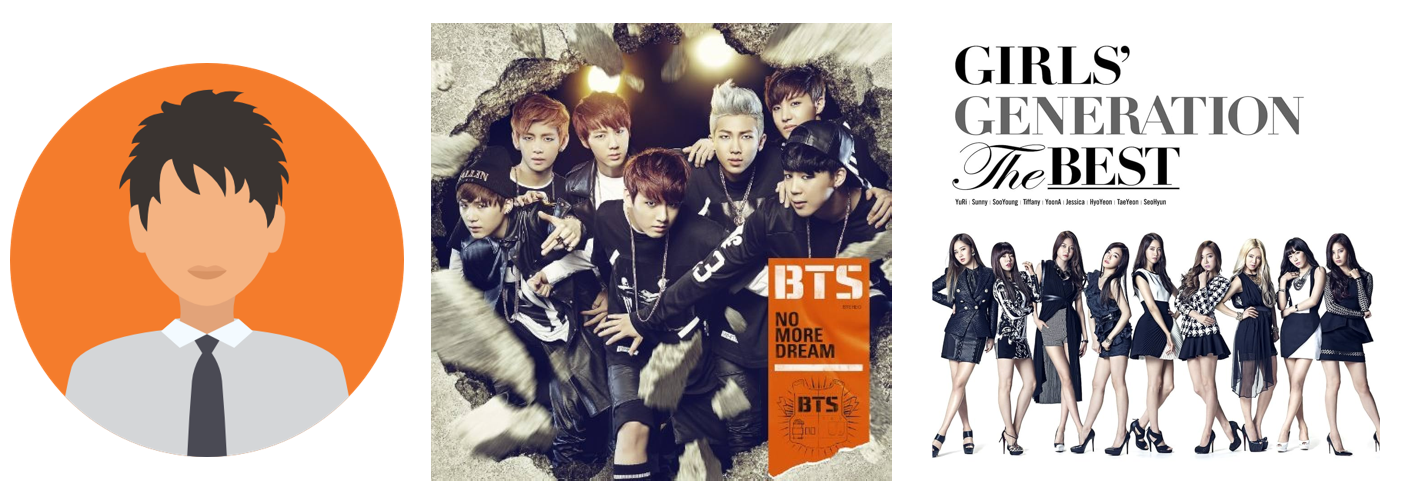}}         &   0.5876             & 11 & \raisebox{-0.7mm}{\includegraphics[width=1.7cm]{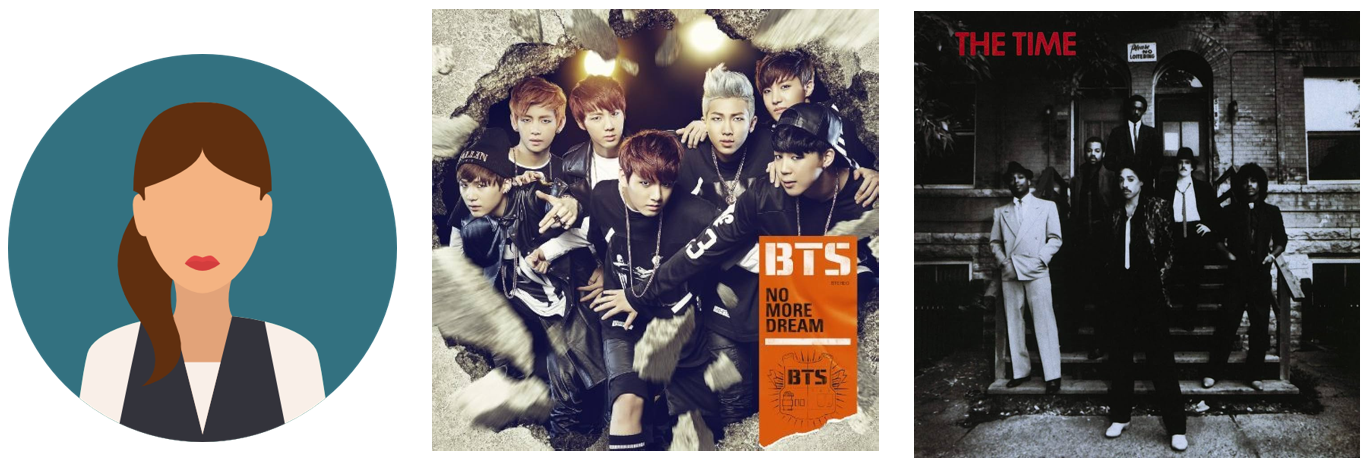}}         &    0.4023             \\ \hline
6 & \raisebox{-0.7mm}{\includegraphics[width=1.7cm]{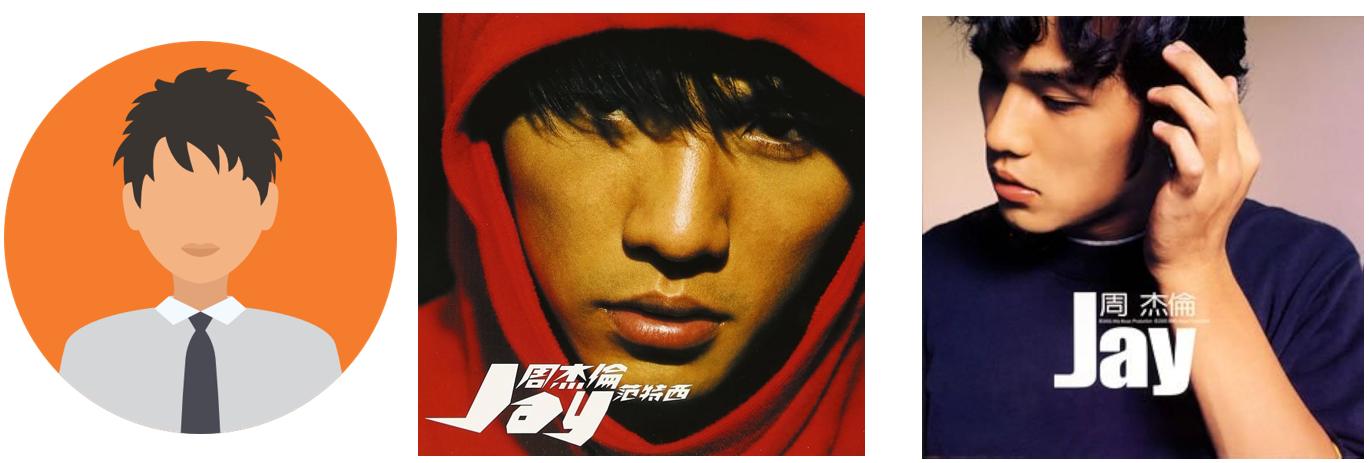}}                               &      -0.3749      & 12     & \raisebox{-0.7mm}{\includegraphics[width=1.7cm]{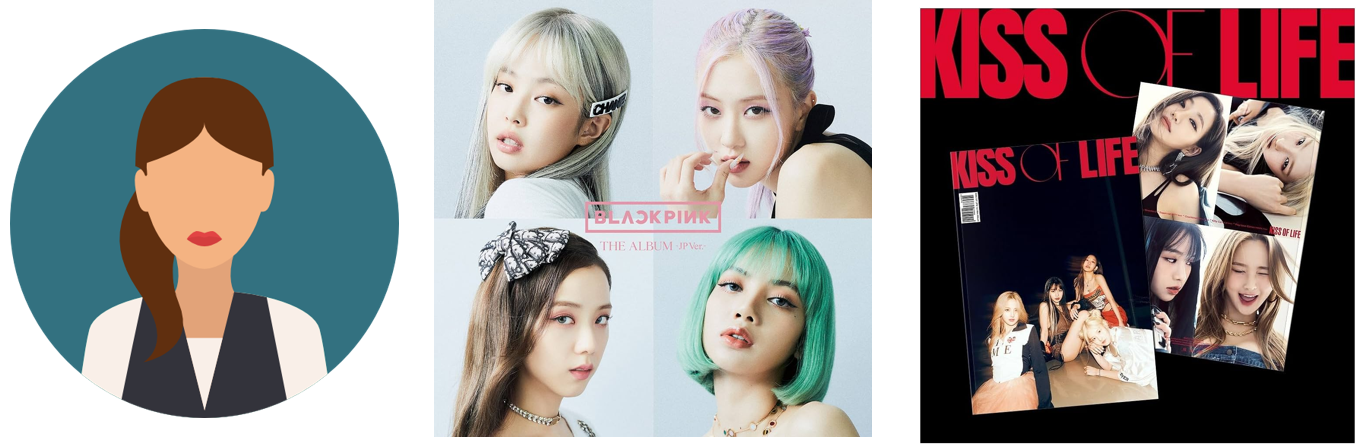}}         &    -0.2178             \\ \hline
\end{tabular}}
\label{table:case study}
\end{table}
\vspace{-1mm}

\subsubsection{\textbf{Stability Analysis (RQ5)}}
\begin{figure}[ht]
    \centering
    \includegraphics[width=0.5\textwidth]{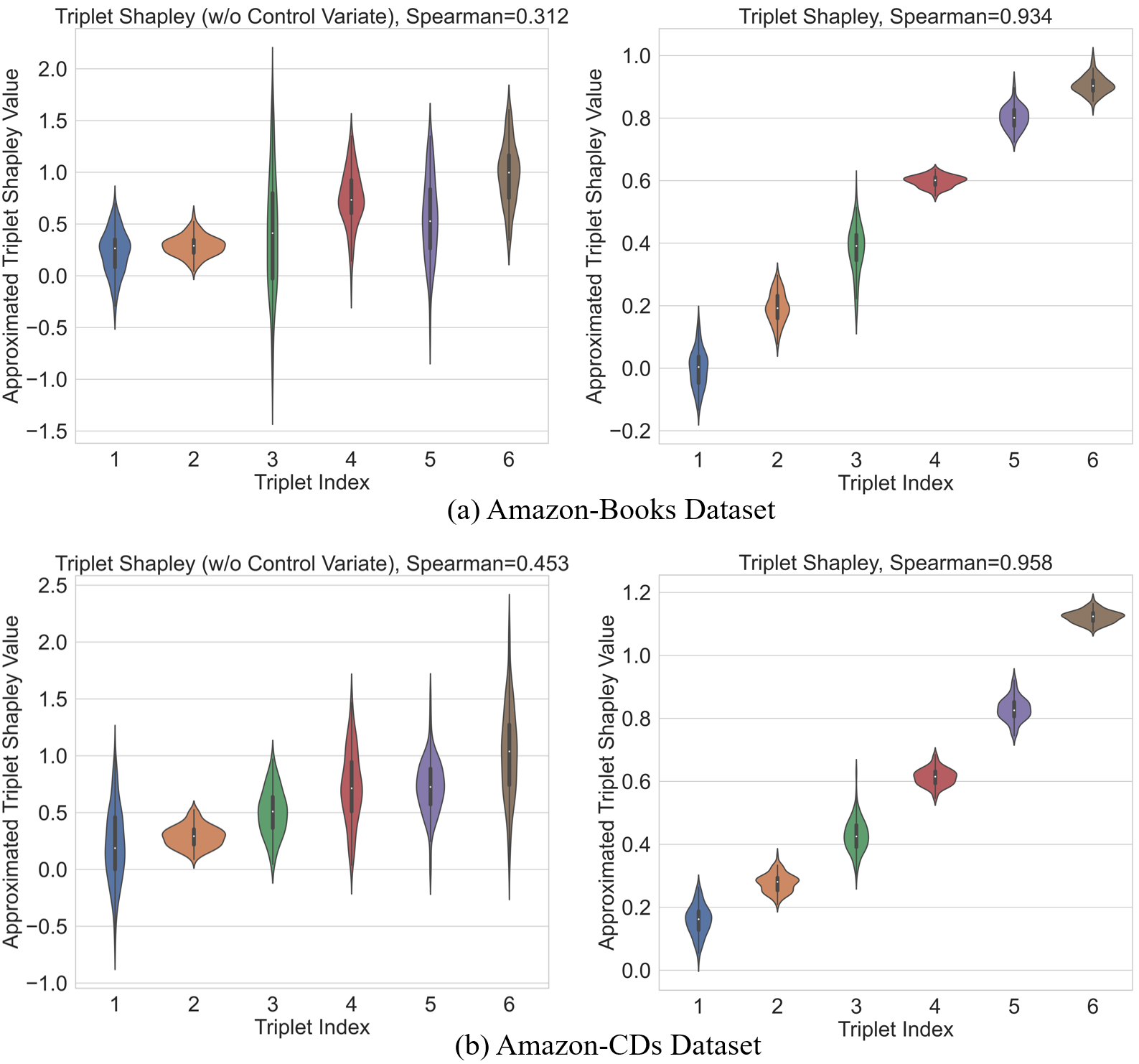}
    \caption{Distribution (in violin plot) of the estimates for triplet Shapley (w/o control variate) and triplet Shapley, based on 6 randomly selected triplets across 5 independent runs on Amazon-Books and Amazon-CDs datasets.
    }
    \label{fig:stability analysis}
    \vspace{-4mm}
\end{figure}

To answer RQ5, We provide the estimated triplet Shapley values of some certain triplets in different trials before and after applying the control variates-based stabilizing strategy. Especially, we use the violin plot and Spearman index to illustrate this comparison in Figure~\ref{fig:stability analysis}.  From this figure, we can first observe that the estimation variance of each sample's triplet Shapley value can be reduced greatly. Second, the rankings among different triplets can be more stable and discriminative in different trials, which is also proved by the Spearman index values. After applying the control variates method, the Spearman index converges to 1, which shows that the rankings obtained via different runs are almost entirely consistent. Thus, our triplet Shapley values are more reproducible and reliable.

\subsubsection{\textbf{Time Complexity (RQ6)}}
\begin{figure}[ht]
    \centering
    \includegraphics[width=0.48\textwidth]{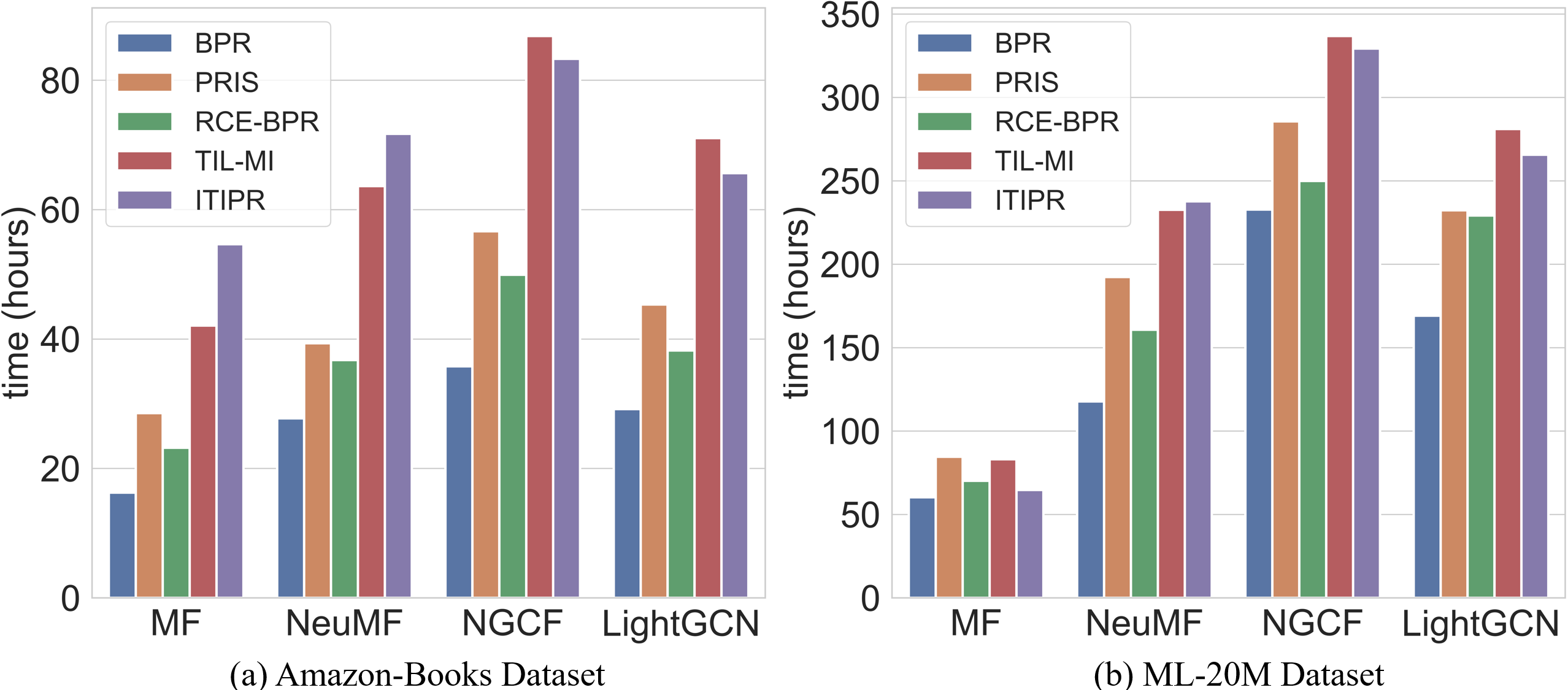}
    \caption{Training time comparison of various methods with different base models on Amazon-Books and ML-20M.
    }
    \label{fig:time complexity}
    \vspace{-0.3cm}
\end{figure}

In Figure~\ref{fig:time complexity}, we illustrate the time consumption of representative methods in each category under different base recommendation models: BPR, TIL-UI, TIL-MI, ITIPR on two datasets with different orders of magnitude data: Amazon-Books and ML-20M datasets, respectively. First, we can observe that \textit{triplet importance modeling} methods generally take more training time than vanilla BPR, \textit{negative item sampling}, and \textit{positive pair re-weighting} approaches, which implies that more time consumption for fine-grained triplet importance modeling is a necessary cost for better model performance. Second, we can find that the time efficiency of our ITIPR is still competitive compared with previous TIL-MI. The corresponding reasons are two folds: 1) our proposed gradient-based truncated Monte Carlo approximation can largely reduce the convergence time; 2) the bi-level optimization in TIL-MI introduces the nested structure and non-convexity issues, which make it hard to converge and computationally expensive.

\section{Conclusion and Future Works}
In this paper, we introduce the triplet Shapley for interpretable triplet importance measurement in Bayesian personalized ranking and improve assessment through a Monte Carlo method with control covariates to reduce variance. Our method, combined with importance-aware resampling, enhances recommendation model training. Experiments on six datasets confirm our approach's effectiveness and efficiency. Future work will focus on efficient computation for large-scale data in practical applications.

\begin{acks}
This work was supported by the Start-up Grant (No. 9610564), the Strategic Research Grant (No. 7005847) of the City University of Hong Kong, and the Early Career Scheme (No. CityU 21219323) of the University Grants Committee (UGC).
\end{acks}

\bibliographystyle{ACM-Reference-Format}
\bibliography{reference.bib}


\begin{thebibliography}{48}


\ifx \showCODEN    \undefined \def \showCODEN     #1{\unskip}     \fi
\ifx \showDOI      \undefined \def \showDOI       #1{#1}\fi
\ifx \showISBNx    \undefined \def \showISBNx     #1{\unskip}     \fi
\ifx \showISBNxiii \undefined \def \showISBNxiii  #1{\unskip}     \fi
\ifx \showISSN     \undefined \def \showISSN      #1{\unskip}     \fi
\ifx \showLCCN     \undefined \def \showLCCN      #1{\unskip}     \fi
\ifx \shownote     \undefined \def \shownote      #1{#1}          \fi
\ifx \showarticletitle \undefined \def \showarticletitle #1{#1}   \fi
\ifx \showURL      \undefined \def \showURL       {\relax}        \fi
\providecommand\bibfield[2]{#2}
\providecommand\bibinfo[2]{#2}
\providecommand\natexlab[1]{#1}
\providecommand\showeprint[2][]{arXiv:#2}

\bibitem[Arora et~al\mbox{.}(2019)]%
        {arora2019implicit}
\bibfield{author}{\bibinfo{person}{Sanjeev Arora}, \bibinfo{person}{Nadav Cohen}, \bibinfo{person}{Wei Hu}, {and} \bibinfo{person}{Yuping Luo}.} \bibinfo{year}{2019}\natexlab{}.
\newblock \showarticletitle{Implicit regularization in deep matrix factorization}.
\newblock \bibinfo{journal}{\emph{Advances in Neural Information Processing Systems}}  \bibinfo{volume}{32} (\bibinfo{year}{2019}).
\newblock


\bibitem[Asghar(2016)]%
        {asghar2016yelp}
\bibfield{author}{\bibinfo{person}{Nabiha Asghar}.} \bibinfo{year}{2016}\natexlab{}.
\newblock \showarticletitle{Yelp dataset challenge: Review rating prediction}.
\newblock \bibinfo{journal}{\emph{arXiv preprint arXiv:1605.05362}} (\bibinfo{year}{2016}).
\newblock


\bibitem[Cho et~al\mbox{.}(2011)]%
        {cho2011friendship}
\bibfield{author}{\bibinfo{person}{Eunjoon Cho}, \bibinfo{person}{Seth~A Myers}, {and} \bibinfo{person}{Jure Leskovec}.} \bibinfo{year}{2011}\natexlab{}.
\newblock \showarticletitle{Friendship and mobility: user movement in location-based social networks}. In \bibinfo{booktitle}{\emph{Proceedings of the 17th ACM SIGKDD international conference on Knowledge discovery and data mining}}. \bibinfo{pages}{1082--1090}.
\newblock


\bibitem[Cook(2000)]%
        {Cook00}
\bibfield{author}{\bibinfo{person}{R.~Dennis Cook}.} \bibinfo{year}{2000}\natexlab{}.
\newblock \showarticletitle{Detection of Influential Observation in Linear Regression}.
\newblock \bibinfo{journal}{\emph{Technometrics}} \bibinfo{volume}{42}, \bibinfo{number}{1} (\bibinfo{year}{2000}), \bibinfo{pages}{65--68}.
\newblock


\bibitem[Cook and Weisberg(1982)]%
        {cook1982residuals}
\bibfield{author}{\bibinfo{person}{R~Dennis Cook} {and} \bibinfo{person}{Sanford Weisberg}.} \bibinfo{year}{1982}\natexlab{}.
\newblock \bibinfo{booktitle}{\emph{Residuals and influence in regression}}.
\newblock \bibinfo{publisher}{New York: Chapman and Hall}.
\newblock


\bibitem[Deshpande and Karypis(2004)]%
        {deshpande2004item}
\bibfield{author}{\bibinfo{person}{Mukund Deshpande} {and} \bibinfo{person}{George Karypis}.} \bibinfo{year}{2004}\natexlab{}.
\newblock \showarticletitle{Item-based top-n recommendation algorithms}.
\newblock \bibinfo{journal}{\emph{ACM Transactions on Information Systems (TOIS)}} \bibinfo{volume}{22}, \bibinfo{number}{1} (\bibinfo{year}{2004}), \bibinfo{pages}{143--177}.
\newblock


\bibitem[Dubey(1975)]%
        {dubey1975uniqueness}
\bibfield{author}{\bibinfo{person}{Pradeep Dubey}.} \bibinfo{year}{1975}\natexlab{}.
\newblock \showarticletitle{On the uniqueness of the Shapley value}.
\newblock \bibinfo{journal}{\emph{International Journal of Game Theory}} \bibinfo{volume}{4}, \bibinfo{number}{3} (\bibinfo{year}{1975}), \bibinfo{pages}{131--139}.
\newblock


\bibitem[Gantner et~al\mbox{.}(2012)]%
        {gantner2012personalized}
\bibfield{author}{\bibinfo{person}{Zeno Gantner}, \bibinfo{person}{Lucas Drumond}, \bibinfo{person}{Christoph Freudenthaler}, {and} \bibinfo{person}{Lars Schmidt-Thieme}.} \bibinfo{year}{2012}\natexlab{}.
\newblock \showarticletitle{Personalized ranking for non-uniformly sampled items}. In \bibinfo{booktitle}{\emph{Proceedings of KDD Cup 2011}}. PMLR, \bibinfo{pages}{231--247}.
\newblock


\bibitem[Ghorbani et~al\mbox{.}(2020)]%
        {disdatavalue}
\bibfield{author}{\bibinfo{person}{Amirata Ghorbani}, \bibinfo{person}{Michael~P. Kim}, {and} \bibinfo{person}{James Zou}.} \bibinfo{year}{2020}\natexlab{}.
\newblock \showarticletitle{A Distributional Framework For Data Valuation}. In \bibinfo{booktitle}{\emph{{ICML}}} \emph{(\bibinfo{series}{Proceedings of Machine Learning Research}, Vol.~\bibinfo{volume}{119})}. \bibinfo{publisher}{{PMLR}}, \bibinfo{pages}{3535--3544}.
\newblock


\bibitem[Ghorbani and Zou(2019)]%
        {datashapley}
\bibfield{author}{\bibinfo{person}{Amirata Ghorbani} {and} \bibinfo{person}{James~Y. Zou}.} \bibinfo{year}{2019}\natexlab{}.
\newblock \showarticletitle{Data Shapley: Equitable Valuation of Data for Machine Learning}. In \bibinfo{booktitle}{\emph{{ICML}}} \emph{(\bibinfo{series}{Proceedings of Machine Learning Research}, Vol.~\bibinfo{volume}{97})}. \bibinfo{publisher}{{PMLR}}, \bibinfo{pages}{2242--2251}.
\newblock


\bibitem[Ghorbani and Zou(2020)]%
        {neuronshapley}
\bibfield{author}{\bibinfo{person}{Amirata Ghorbani} {and} \bibinfo{person}{James~Y. Zou}.} \bibinfo{year}{2020}\natexlab{}.
\newblock \showarticletitle{Neuron Shapley: Discovering the Responsible Neurons}. In \bibinfo{booktitle}{\emph{NeurIPS}}.
\newblock


\bibitem[Harper and Konstan(2015)]%
        {harper2015movielens}
\bibfield{author}{\bibinfo{person}{F~Maxwell Harper} {and} \bibinfo{person}{Joseph~A Konstan}.} \bibinfo{year}{2015}\natexlab{}.
\newblock \showarticletitle{The movielens datasets: History and context}.
\newblock \bibinfo{journal}{\emph{Acm transactions on interactive intelligent systems (tiis)}} \bibinfo{volume}{5}, \bibinfo{number}{4} (\bibinfo{year}{2015}), \bibinfo{pages}{1--19}.
\newblock


\bibitem[He et~al\mbox{.}(2023b)]%
        {he2023dynamic}
\bibfield{author}{\bibinfo{person}{Bowei He}, \bibinfo{person}{Xu He}, \bibinfo{person}{Renrui Zhang}, \bibinfo{person}{Yingxue Zhang}, \bibinfo{person}{Ruiming Tang}, {and} \bibinfo{person}{Chen Ma}.} \bibinfo{year}{2023}\natexlab{b}.
\newblock \showarticletitle{Dynamic Embedding Size Search with Minimum Regret for Streaming Recommender System}. In \bibinfo{booktitle}{\emph{Proceedings of the 32nd ACM International Conference on Information and Knowledge Management}}. \bibinfo{pages}{741--750}.
\newblock


\bibitem[He et~al\mbox{.}(2023a)]%
        {he2023dynamically}
\bibfield{author}{\bibinfo{person}{Bowei He}, \bibinfo{person}{Xu He}, \bibinfo{person}{Yingxue Zhang}, \bibinfo{person}{Ruiming Tang}, {and} \bibinfo{person}{Chen Ma}.} \bibinfo{year}{2023}\natexlab{a}.
\newblock \showarticletitle{Dynamically expandable graph convolution for streaming recommendation}. In \bibinfo{booktitle}{\emph{Proceedings of the ACM Web Conference 2023}}. \bibinfo{pages}{1457--1467}.
\newblock


\bibitem[He and McAuley(2016)]%
        {he2016ups}
\bibfield{author}{\bibinfo{person}{Ruining He} {and} \bibinfo{person}{Julian McAuley}.} \bibinfo{year}{2016}\natexlab{}.
\newblock \showarticletitle{Ups and downs: Modeling the visual evolution of fashion trends with one-class collaborative filtering}. In \bibinfo{booktitle}{\emph{proceedings of the 25th international conference on world wide web}}. \bibinfo{pages}{507--517}.
\newblock


\bibitem[He et~al\mbox{.}(2020)]%
        {he2020lightgcn}
\bibfield{author}{\bibinfo{person}{Xiangnan He}, \bibinfo{person}{Kuan Deng}, \bibinfo{person}{Xiang Wang}, \bibinfo{person}{Yan Li}, \bibinfo{person}{Yongdong Zhang}, {and} \bibinfo{person}{Meng Wang}.} \bibinfo{year}{2020}\natexlab{}.
\newblock \showarticletitle{Lightgcn: Simplifying and powering graph convolution network for recommendation}. In \bibinfo{booktitle}{\emph{Proceedings of the 43rd International ACM SIGIR conference on research and development in Information Retrieval}}. \bibinfo{pages}{639--648}.
\newblock


\bibitem[He et~al\mbox{.}(2017)]%
        {he2017neural}
\bibfield{author}{\bibinfo{person}{Xiangnan He}, \bibinfo{person}{Lizi Liao}, \bibinfo{person}{Hanwang Zhang}, \bibinfo{person}{Liqiang Nie}, \bibinfo{person}{Xia Hu}, {and} \bibinfo{person}{Tat-Seng Chua}.} \bibinfo{year}{2017}\natexlab{}.
\newblock \showarticletitle{Neural collaborative filtering}. In \bibinfo{booktitle}{\emph{Proceedings of the 26th international conference on world wide web}}. \bibinfo{pages}{173--182}.
\newblock


\bibitem[Horv{\'a}th et~al\mbox{.}(2023)]%
        {horvath2023stochastic}
\bibfield{author}{\bibinfo{person}{Samuel Horv{\'a}th}, \bibinfo{person}{Dmitry Kovalev}, \bibinfo{person}{Konstantin Mishchenko}, \bibinfo{person}{Peter Richt{\'a}rik}, {and} \bibinfo{person}{Sebastian Stich}.} \bibinfo{year}{2023}\natexlab{}.
\newblock \showarticletitle{Stochastic distributed learning with gradient quantization and double-variance reduction}.
\newblock \bibinfo{journal}{\emph{Optimization Methods and Software}} \bibinfo{volume}{38}, \bibinfo{number}{1} (\bibinfo{year}{2023}), \bibinfo{pages}{91--106}.
\newblock


\bibitem[Hossain and Naik(1989)]%
        {hossain1989detection}
\bibfield{author}{\bibinfo{person}{A Hossain} {and} \bibinfo{person}{DN Naik}.} \bibinfo{year}{1989}\natexlab{}.
\newblock \showarticletitle{Detection of influential observations in multivariate regression}.
\newblock \bibinfo{journal}{\emph{Journal of Applied Statistics}} \bibinfo{volume}{16}, \bibinfo{number}{1} (\bibinfo{year}{1989}), \bibinfo{pages}{25--37}.
\newblock


\bibitem[Jethani et~al\mbox{.}(2021)]%
        {jethani2021fastshap}
\bibfield{author}{\bibinfo{person}{Neil Jethani}, \bibinfo{person}{Mukund Sudarshan}, \bibinfo{person}{Ian~Connick Covert}, \bibinfo{person}{Su-In Lee}, {and} \bibinfo{person}{Rajesh Ranganath}.} \bibinfo{year}{2021}\natexlab{}.
\newblock \showarticletitle{FastSHAP: Real-Time Shapley Value Estimation}. In \bibinfo{booktitle}{\emph{International Conference on Learning Representations}}.
\newblock


\bibitem[Jia et~al\mbox{.}(2019)]%
        {JiaDWHGLZSS19}
\bibfield{author}{\bibinfo{person}{Ruoxi Jia}, \bibinfo{person}{David Dao}, \bibinfo{person}{Boxin Wang}, \bibinfo{person}{Frances~Ann Hubis}, \bibinfo{person}{Nezihe~Merve G{\"{u}}rel}, \bibinfo{person}{Bo Li}, \bibinfo{person}{Ce Zhang}, \bibinfo{person}{Costas~J. Spanos}, {and} \bibinfo{person}{Dawn Song}.} \bibinfo{year}{2019}\natexlab{}.
\newblock \showarticletitle{Efficient Task-Specific Data Valuation for Nearest Neighbor Algorithms}.
\newblock \bibinfo{journal}{\emph{Proc. {VLDB} Endow.}} \bibinfo{volume}{12}, \bibinfo{number}{11} (\bibinfo{year}{2019}), \bibinfo{pages}{1610--1623}.
\newblock


\bibitem[Karimireddy et~al\mbox{.}(2020)]%
        {karimireddy2020scaffold}
\bibfield{author}{\bibinfo{person}{Sai~Praneeth Karimireddy}, \bibinfo{person}{Satyen Kale}, \bibinfo{person}{Mehryar Mohri}, \bibinfo{person}{Sashank Reddi}, \bibinfo{person}{Sebastian Stich}, {and} \bibinfo{person}{Ananda~Theertha Suresh}.} \bibinfo{year}{2020}\natexlab{}.
\newblock \showarticletitle{Scaffold: Stochastic controlled averaging for federated learning}. In \bibinfo{booktitle}{\emph{International conference on machine learning}}. PMLR, \bibinfo{pages}{5132--5143}.
\newblock


\bibitem[Karp et~al\mbox{.}(1989)]%
        {karp1989monte}
\bibfield{author}{\bibinfo{person}{Richard~M Karp}, \bibinfo{person}{Michael Luby}, {and} \bibinfo{person}{Neal Madras}.} \bibinfo{year}{1989}\natexlab{}.
\newblock \showarticletitle{Monte-Carlo approximation algorithms for enumeration problems}.
\newblock \bibinfo{journal}{\emph{Journal of algorithms}} \bibinfo{volume}{10}, \bibinfo{number}{3} (\bibinfo{year}{1989}), \bibinfo{pages}{429--448}.
\newblock


\bibitem[Kim et~al\mbox{.}(2014)]%
        {kim2014modeling}
\bibfield{author}{\bibinfo{person}{Youngho Kim}, \bibinfo{person}{Ahmed Hassan}, \bibinfo{person}{Ryen~W White}, {and} \bibinfo{person}{Imed Zitouni}.} \bibinfo{year}{2014}\natexlab{}.
\newblock \showarticletitle{Modeling dwell time to predict click-level satisfaction}. In \bibinfo{booktitle}{\emph{Proceedings of the 7th ACM international conference on Web search and data mining}}. \bibinfo{pages}{193--202}.
\newblock


\bibitem[Kipf and Welling(2017)]%
        {max2017gcn}
\bibfield{author}{\bibinfo{person}{Thomas~N. Kipf} {and} \bibinfo{person}{Max Welling}.} \bibinfo{year}{2017}\natexlab{}.
\newblock \showarticletitle{Semi-Supervised Classification with Graph Convolutional Networks}. In \bibinfo{booktitle}{\emph{{ICLR} (Poster)}}. \bibinfo{publisher}{OpenReview.net}.
\newblock


\bibitem[Koren et~al\mbox{.}(2009)]%
        {koren2009matrix}
\bibfield{author}{\bibinfo{person}{Yehuda Koren}, \bibinfo{person}{Robert Bell}, {and} \bibinfo{person}{Chris Volinsky}.} \bibinfo{year}{2009}\natexlab{}.
\newblock \showarticletitle{Matrix factorization techniques for recommender systems}.
\newblock \bibinfo{journal}{\emph{Computer}} \bibinfo{volume}{42}, \bibinfo{number}{8} (\bibinfo{year}{2009}), \bibinfo{pages}{30--37}.
\newblock


\bibitem[Kwon et~al\mbox{.}(2021)]%
        {disshapley}
\bibfield{author}{\bibinfo{person}{Yongchan Kwon}, \bibinfo{person}{Manuel~A. Rivas}, {and} \bibinfo{person}{James Zou}.} \bibinfo{year}{2021}\natexlab{}.
\newblock \showarticletitle{Efficient Computation and Analysis of Distributional Shapley Values}. In \bibinfo{booktitle}{\emph{{AISTATS}}} \emph{(\bibinfo{series}{Proceedings of Machine Learning Research}, Vol.~\bibinfo{volume}{130})}. \bibinfo{publisher}{{PMLR}}, \bibinfo{pages}{793--801}.
\newblock


\bibitem[Kwon and Zou(2022)]%
        {betashapley}
\bibfield{author}{\bibinfo{person}{Yongchan Kwon} {and} \bibinfo{person}{James Zou}.} \bibinfo{year}{2022}\natexlab{}.
\newblock \showarticletitle{Beta Shapley: a Unified and Noise-reduced Data Valuation Framework for Machine Learning}. In \bibinfo{booktitle}{\emph{{AISTATS}}} \emph{(\bibinfo{series}{Proceedings of Machine Learning Research}, Vol.~\bibinfo{volume}{151})}. \bibinfo{publisher}{{PMLR}}, \bibinfo{pages}{8780--8802}.
\newblock


\bibitem[Lian et~al\mbox{.}(2020)]%
        {lian2020personalized}
\bibfield{author}{\bibinfo{person}{Defu Lian}, \bibinfo{person}{Qi Liu}, {and} \bibinfo{person}{Enhong Chen}.} \bibinfo{year}{2020}\natexlab{}.
\newblock \showarticletitle{Personalized ranking with importance sampling}. In \bibinfo{booktitle}{\emph{Proceedings of The Web Conference 2020}}. \bibinfo{pages}{1093--1103}.
\newblock


\bibitem[Lu et~al\mbox{.}(2018)]%
        {lu2018between}
\bibfield{author}{\bibinfo{person}{Hongyu Lu}, \bibinfo{person}{Min Zhang}, {and} \bibinfo{person}{Shaoping Ma}.} \bibinfo{year}{2018}\natexlab{}.
\newblock \showarticletitle{Between clicks and satisfaction: Study on multi-phase user preferences and satisfaction for online news reading}. In \bibinfo{booktitle}{\emph{The 41st International ACM SIGIR Conference on Research \& Development in Information Retrieval}}. \bibinfo{pages}{435--444}.
\newblock


\bibitem[Mohamed et~al\mbox{.}(2020)]%
        {mohamed2020monte}
\bibfield{author}{\bibinfo{person}{Shakir Mohamed}, \bibinfo{person}{Mihaela Rosca}, \bibinfo{person}{Michael Figurnov}, {and} \bibinfo{person}{Andriy Mnih}.} \bibinfo{year}{2020}\natexlab{}.
\newblock \showarticletitle{Monte carlo gradient estimation in machine learning}.
\newblock \bibinfo{journal}{\emph{The Journal of Machine Learning Research}} \bibinfo{volume}{21}, \bibinfo{number}{1} (\bibinfo{year}{2020}), \bibinfo{pages}{5183--5244}.
\newblock


\bibitem[Rendle and Freudenthaler(2014)]%
        {rendle2014improving}
\bibfield{author}{\bibinfo{person}{Steffen Rendle} {and} \bibinfo{person}{Christoph Freudenthaler}.} \bibinfo{year}{2014}\natexlab{}.
\newblock \showarticletitle{Improving pairwise learning for item recommendation from implicit feedback}. In \bibinfo{booktitle}{\emph{Proceedings of the 7th ACM international conference on Web search and data mining}}. \bibinfo{pages}{273--282}.
\newblock


\bibitem[Rendle et~al\mbox{.}(2009)]%
        {bpr2008}
\bibfield{author}{\bibinfo{person}{Steffen Rendle}, \bibinfo{person}{Christoph Freudenthaler}, \bibinfo{person}{Zeno Gantner}, {and} \bibinfo{person}{Lars Schmidt{-}Thieme}.} \bibinfo{year}{2009}\natexlab{}.
\newblock \showarticletitle{{BPR:} Bayesian Personalized Ranking from Implicit Feedback}. In \bibinfo{booktitle}{\emph{{UAI} 2009, Proceedings of the Twenty-Fifth Conference on Uncertainty in Artificial Intelligence, Montreal, QC, Canada, June 18-21, 2009}}. \bibinfo{publisher}{{AUAI} Press}.
\newblock


\bibitem[Roth(1988)]%
        {roth1988shapley}
\bibfield{author}{\bibinfo{person}{Alvin~E Roth}.} \bibinfo{year}{1988}\natexlab{}.
\newblock \bibinfo{booktitle}{\emph{The Shapley value: essays in honor of Lloyd S. Shapley}}.
\newblock \bibinfo{publisher}{Cambridge University Press}.
\newblock


\bibitem[Salakhutdinov et~al\mbox{.}(2007)]%
        {salakhutdinov2007restricted}
\bibfield{author}{\bibinfo{person}{Ruslan Salakhutdinov}, \bibinfo{person}{Andriy Mnih}, {and} \bibinfo{person}{Geoffrey Hinton}.} \bibinfo{year}{2007}\natexlab{}.
\newblock \showarticletitle{Restricted Boltzmann machines for collaborative filtering}. In \bibinfo{booktitle}{\emph{Proceedings of the 24th international conference on Machine learning}}. \bibinfo{pages}{791--798}.
\newblock


\bibitem[Sarwar et~al\mbox{.}(2001)]%
        {sarwar2001item}
\bibfield{author}{\bibinfo{person}{Badrul Sarwar}, \bibinfo{person}{George Karypis}, \bibinfo{person}{Joseph Konstan}, {and} \bibinfo{person}{John Riedl}.} \bibinfo{year}{2001}\natexlab{}.
\newblock \showarticletitle{Item-based collaborative filtering recommendation algorithms}. In \bibinfo{booktitle}{\emph{Proceedings of the 10th international conference on World Wide Web}}. \bibinfo{pages}{285--295}.
\newblock


\bibitem[Schoch et~al\mbox{.}(2022)]%
        {schoch2022cs}
\bibfield{author}{\bibinfo{person}{Stephanie Schoch}, \bibinfo{person}{Haifeng Xu}, {and} \bibinfo{person}{Yangfeng Ji}.} \bibinfo{year}{2022}\natexlab{}.
\newblock \showarticletitle{CS-Shapley: Class-wise Shapley Values for Data Valuation in Classification}.
\newblock \bibinfo{journal}{\emph{arXiv preprint arXiv:2211.06800}} (\bibinfo{year}{2022}).
\newblock


\bibitem[Sun et~al\mbox{.}(2019)]%
        {sun2019multi}
\bibfield{author}{\bibinfo{person}{Jianing Sun}, \bibinfo{person}{Yingxue Zhang}, \bibinfo{person}{Chen Ma}, \bibinfo{person}{Mark Coates}, \bibinfo{person}{Huifeng Guo}, \bibinfo{person}{Ruiming Tang}, {and} \bibinfo{person}{Xiuqiang He}.} \bibinfo{year}{2019}\natexlab{}.
\newblock \showarticletitle{Multi-graph convolution collaborative filtering}. In \bibinfo{booktitle}{\emph{2019 IEEE international conference on data mining (ICDM)}}. IEEE, \bibinfo{pages}{1306--1311}.
\newblock


\bibitem[Sun et~al\mbox{.}(2023)]%
        {sun2023vector}
\bibfield{author}{\bibinfo{person}{Zhuo Sun}, \bibinfo{person}{Alessandro Barp}, {and} \bibinfo{person}{Fran{\c{c}}ois-Xavier Briol}.} \bibinfo{year}{2023}\natexlab{}.
\newblock \showarticletitle{Vector-valued control variates}. In \bibinfo{booktitle}{\emph{International Conference on Machine Learning}}. PMLR, \bibinfo{pages}{32819--32846}.
\newblock


\bibitem[van~den Berg et~al\mbox{.}(2017)]%
        {max2017gcmc}
\bibfield{author}{\bibinfo{person}{Rianne van~den Berg}, \bibinfo{person}{Thomas~N. Kipf}, {and} \bibinfo{person}{Max Welling}.} \bibinfo{year}{2017}\natexlab{}.
\newblock \showarticletitle{Graph Convolutional Matrix Completion}.
\newblock \bibinfo{journal}{\emph{CoRR}}  \bibinfo{volume}{abs/1706.02263} (\bibinfo{year}{2017}).
\newblock


\bibitem[Wang et~al\mbox{.}(2021)]%
        {wang2021denoising}
\bibfield{author}{\bibinfo{person}{Wenjie Wang}, \bibinfo{person}{Fuli Feng}, \bibinfo{person}{Xiangnan He}, \bibinfo{person}{Liqiang Nie}, {and} \bibinfo{person}{Tat-Seng Chua}.} \bibinfo{year}{2021}\natexlab{}.
\newblock \showarticletitle{Denoising implicit feedback for recommendation}. In \bibinfo{booktitle}{\emph{Proceedings of the 14th ACM international conference on web search and data mining}}. \bibinfo{pages}{373--381}.
\newblock


\bibitem[Wang et~al\mbox{.}(2019)]%
        {wang2019neural}
\bibfield{author}{\bibinfo{person}{Xiang Wang}, \bibinfo{person}{Xiangnan He}, \bibinfo{person}{Meng Wang}, \bibinfo{person}{Fuli Feng}, {and} \bibinfo{person}{Tat-Seng Chua}.} \bibinfo{year}{2019}\natexlab{}.
\newblock \showarticletitle{Neural graph collaborative filtering}. In \bibinfo{booktitle}{\emph{Proceedings of the 42nd international ACM SIGIR conference on Research and development in Information Retrieval}}. \bibinfo{pages}{165--174}.
\newblock


\bibitem[Wen et~al\mbox{.}(2019)]%
        {wen2019leveraging}
\bibfield{author}{\bibinfo{person}{Hongyi Wen}, \bibinfo{person}{Longqi Yang}, {and} \bibinfo{person}{Deborah Estrin}.} \bibinfo{year}{2019}\natexlab{}.
\newblock \showarticletitle{Leveraging post-click feedback for content recommendations}. In \bibinfo{booktitle}{\emph{Proceedings of the 13th ACM Conference on Recommender Systems}}. \bibinfo{pages}{278--286}.
\newblock


\bibitem[Wu et~al\mbox{.}(2022)]%
        {wu2022adapting}
\bibfield{author}{\bibinfo{person}{Haolun Wu}, \bibinfo{person}{Chen Ma}, \bibinfo{person}{Yingxue Zhang}, \bibinfo{person}{Xue Liu}, \bibinfo{person}{Ruiming Tang}, {and} \bibinfo{person}{Mark Coates}.} \bibinfo{year}{2022}\natexlab{}.
\newblock \showarticletitle{Adapting Triplet Importance of Implicit Feedback for Personalized Recommendation}. In \bibinfo{booktitle}{\emph{Proceedings of the 31st ACM International Conference on Information \& Knowledge Management}}. \bibinfo{pages}{2148--2157}.
\newblock


\bibitem[Xie et~al\mbox{.}(2022)]%
        {xie2022reweighting}
\bibfield{author}{\bibinfo{person}{Ruobing Xie}, \bibinfo{person}{Lin Ma}, \bibinfo{person}{Shaoliang Zhang}, \bibinfo{person}{Feng Xia}, {and} \bibinfo{person}{Leyu Lin}.} \bibinfo{year}{2022}\natexlab{}.
\newblock \showarticletitle{Reweighting Clicks with Dwell Time in Recommendation}.
\newblock \bibinfo{journal}{\emph{arXiv preprint arXiv:2209.09000}} (\bibinfo{year}{2022}).
\newblock


\bibitem[Yan and Procaccia(2021)]%
        {yan2021if}
\bibfield{author}{\bibinfo{person}{Tom Yan} {and} \bibinfo{person}{Ariel~D Procaccia}.} \bibinfo{year}{2021}\natexlab{}.
\newblock \showarticletitle{If you like shapley then you’ll love the core}. In \bibinfo{booktitle}{\emph{Proceedings of the AAAI Conference on Artificial Intelligence}}, Vol.~\bibinfo{volume}{35}. \bibinfo{pages}{5751--5759}.
\newblock


\bibitem[Yi et~al\mbox{.}(2014)]%
        {yi2014beyond}
\bibfield{author}{\bibinfo{person}{Xing Yi}, \bibinfo{person}{Liangjie Hong}, \bibinfo{person}{Erheng Zhong}, \bibinfo{person}{Nanthan~Nan Liu}, {and} \bibinfo{person}{Suju Rajan}.} \bibinfo{year}{2014}\natexlab{}.
\newblock \showarticletitle{Beyond clicks: dwell time for personalization}. In \bibinfo{booktitle}{\emph{Proceedings of the 8th ACM Conference on Recommender systems}}. \bibinfo{pages}{113--120}.
\newblock


\bibitem[Zhang et~al\mbox{.}(2022)]%
        {zhang2022towards}
\bibfield{author}{\bibinfo{person}{Zhao-Yu Zhang}, \bibinfo{person}{Xiang-Rong Sheng}, \bibinfo{person}{Yujing Zhang}, \bibinfo{person}{Biye Jiang}, \bibinfo{person}{Shuguang Han}, \bibinfo{person}{Hongbo Deng}, {and} \bibinfo{person}{Bo Zheng}.} \bibinfo{year}{2022}\natexlab{}.
\newblock \showarticletitle{Towards Understanding the Overfitting Phenomenon of Deep Click-Through Rate Models}. In \bibinfo{booktitle}{\emph{Proceedings of the 31st ACM International Conference on Information \& Knowledge Management}}. \bibinfo{pages}{2671--2680}.
\newblock


\end{thebibliography}

\newpage
\appendix
\section{Supplementary Preliminaries}
\subsection{Main Notations}
The main notations in this paper are summarized in Table~\ref{table:notation}.
\begin{table}[]
    \caption{Summary of main notations in this work.}
    \centering
    \renewcommand{\arraystretch}{1.3}
    \resizebox{\linewidth}{!}{
    \begin{tabular}{cl}
    \toprule
      {\bfseries Symbol} & {\bfseries Description}\\
      \midrule
      $\mathcal{U}, \mathcal{V}, \mathcal{I}$ & The set of users, items, and user-item interactions \\
      $\mathbf{p}_u, \mathbf{q}_i$ & The user $u$'s embedding and the item $i$'s embedding\\
      $D_S, D^T$ & The whole triplet construction space and the sampled triplet set\\
      $W^{TI}_{(u,i,j)}$ & The triplet importance weight for the triplet $(u,i,j)$ in BPR learning\\
      $V^{TS}_{(u,i,j)}$ & The triplet Shapley for the triplet $(u, i, j) \in D^{T}$\\
      $V^{TS, *}_{(u,i,j)}$ & The control variate that is known and correlated with $V^{TS}_{(u,i,j)}$ \\
      $\Pi^T$ & The set of all possible permutations consisting of the triplets in $D^{T}$ \\
      $P^{\pi}_{(u,i,j)}$ & The set of triplets that precede $(u,i,j)$ in the permutation $\pi \in \Pi^{T}$\\
      $F^{\pi}_{(u,i,j)}$ & The set of triplets that follow $(u,i,j)$ in the permutation $\pi \in \Pi^{T}$\\
      $G(\cdot)$ & The group with all possible permutations consisted of same triplets \\
      $\pi_k[s]$ & The $s$-th triplet in the permutation from the $k$-th sampling\\
      \bottomrule
    \end{tabular}
    }
\label{table:notation}
\end{table}

\subsection{Base Recommendation Models}
\label{sec: rec models}
Matrix factorization (MF)-based models~\citep{koren2009matrix, he2017neural, arora2019implicit} and Graph Neural Network (GNN)-based models~\citep{wang2019neural, he2020lightgcn, sun2019multi, max2017gcn, max2017gcmc} are two categories of most representative and commonly used recommendation models. In this paper, we adopt some popular and effective instances of them to demonstrate the applicability and robustness of our proposed framework. We provide the brief introduction as follows.

\textbf{MF-based Models:} MF~\citep{koren2009matrix} is a pioneer and fundamental method in this line of research. In MF, the user-item interaction matrix is factorized into two matrices that represent the embeddings of users and items. The inner product of user $u$'s embedding $\mathbf{p}_u$ and item $i$'s embedding $\mathbf{q}_i$ is used to predict the user-item interaction between them:
\begin{equation}
        \hat{y}_{u,i} = \mathbf{p}_u \cdot {\mathbf{q}_i}^{\top}.
\end{equation}
The goal of MF is to minimize the difference between the predicted and actual ratings.
Neural Matrix Factorization (NeuMF)~\citep{he2017neural} is an extension of MF that uses neural networks to learn embeddings of users and items, which allows more complex interactions between them. The output of the neural network which utilizes the $u$'s embedding $\mathbf{p}_u$ and item $i$'s embedding $\mathbf{q}_i$ as the input is used to predict the user-item interaction between them:
\begin{equation}
        \hat{y}_{u,i} = MLP(\mathbf{p}_u \odot {\mathbf{q}_i}, MLP(\mathbf{p}_u, {\mathbf{q}_i})),
\end{equation}
where $MLP$ represents the multi-layer perceptron. NMF has been shown to outperform traditional MF in terms of prediction accuracy and scalability. 

\textbf{GNN-based Models:} Neural Graph Collaborative Filtering (NGCF)\\~\citep{wang2019neural} is a graph-based recommender system model that leverages the power of GNN to learn user and item embeddings. NGCF considers the user-item interactions as a bipartite graph and uses GNN to propagate information between connected nodes in the graph. The propagation is formulated as follows:
\begin{equation}
\begin{aligned}
        \mathbf{p}^{(l)}_u &= \sigma(\mathbf{W}_1 \mathbf{p}^{(l-1)}_u + \underset{i \in \mathcal{N}_u}{\sum} \frac{1}{\sqrt{\vert \mathcal{N}_u \vert \vert \mathcal{N}_i \vert}} (\mathbf{W}_1 \mathbf{q}^{(l-1)}_i \\ &+  \mathbf{W}_2 (\mathbf{q}^{(l-1)}_i \odot \mathbf{p}^{(l-1)}_u))),\\
        \mathbf{q}^{(l)}_i &= \sigma(\mathbf{W}_1 \mathbf{q}^{(l-1)}_i + \underset{u \in \mathcal{N}_i}{\sum} \frac{1}{\sqrt{\vert \mathcal{N}_u \vert \vert \mathcal{N}_i \vert}} (\mathbf{W}_1 \mathbf{p}^{(l-1)}_u \\ &+  \mathbf{W}_2 (\mathbf{p}^{(l-1)}_u \odot \mathbf{q}^{(l-1)}_i))),        
\end{aligned}
\end{equation}
where $\mathbf{p}^{(l)}_u$ and $\mathbf{q}^{(l)}_i$ denote the refined embedding of user $u$ and item $i$ after $l$ layers propagation, respectively. $\mathbf{p}^0_u = \mathbf{p}_u, \mathbf{q}^0_i = \mathbf{q}_i$, $\mathcal{N}_u$ denotes the set of items that are interacted by user $u$, and $\mathcal{N}_i$ denotes the set of users that interact with item $i$. $\mathbf{W}_1$ and $\mathbf{W}_2$ are learnable feature transformation matrices. $\sigma(\cdot)$ is the nonlinear activation function, like sigmoid function. After propagation with $L$ layers, the learned embeddings are used to predict the user-item interactions:
\begin{equation}
        \hat{y}_{u,i} = (\mathbf{p}^{(0)}_u || \cdots || \mathbf{p}^{(L)}_u)(\mathbf{q}^{(0)}_i || \cdots ||\mathbf{q}^{(L)}_i)^\top.
\end{equation}
where $||$ is the concatenation operation. NGCF has shown to outperform traditional collaborative filtering methods in accuracy. LightGCN~\citep{he2020lightgcn} is a simplified version of NGCF that only uses the GNN layers without any nonlinear activation functions or additional weight matrices. By reducing the layer numbers and using a simpler aggregation function, LightGCN aims to address the over-smoothing problem that occur in the GNN-based methods. Besides, LightGCN leverages a normalized adjacency matrix to prevent the model from being biased towards highly-connected nodes in the graph. The propagation process in LightGCN is as follows:
\begin{equation}
\begin{aligned}
        \mathbf{p}^{(l)}_u = \underset{i \in \mathcal{N}_u}{\sum} \frac{1}{\sqrt{\vert \mathcal{N}_u \vert \vert \mathcal{N}_i \vert}} \mathbf{q}^{(l-1)}_i,\\
        \mathbf{q}^{(l)}_i = \underset{u \in \mathcal{N}_i}{\sum} \frac{1}{\sqrt{\vert \mathcal{N}_i \vert \vert \mathcal{N}_u \vert}} \mathbf{p}^{(l-1)}_u.
\end{aligned}
\end{equation}
After $L$ times of message propagation, the model prediction is formulated as: 
\begin{equation}
        \hat{y}_{u,i} = \left(\frac{1}{L+1} \underset{l=0}{\sum^L}\mathbf{p}^{(l)}_u\right) \cdot {\left(\frac{1}{L+1} \underset{l=0}{\sum^L} \mathbf{q}^{(l)}_i\right)}^{\mathrm{\top}}.
\end{equation}
It should be noted that LightGCN has shown competitive results compared with some more complex graph-based recommendation models despite its simplicity in structure.

\section{Supplementary Methodology}
\subsection{Proof for Theorem 1}
We provide the following proof for the Theorem~\ref{theo1}.
\vspace{-3mm}
\begin{proof}
First, recall Eq.~\ref{equ:definition} and set $C = \frac{1}{|D^{T}|}$. Then, the following transformation holds:

\begin{equation}
\begin{aligned}
    & V^{TS}_{(u,i,j)} = \frac{1}{|D^{T}|} \underset{S \subset \{(u,i,j)\}^{'}}{\sum} \frac{A(S \cup \left\{(u,i,j)\}\right) - A(S)}{\binom{|D^{T}|-1}{|S|}}     \\
    &= \underset{S \subset \{(u,i,j)\}^{'}}{\sum} \frac{|S|!(|D^{T}|-|S|-1)!}{|D^{T}|!} [A(S \cup \left\{(u,i,j)\}\right) - A(S)] \\
    &= \frac{1}{|D^{T}|} \sum^{|D^{T}|-1}_{k=0} \underset{\substack{|S|=k\\ (u,i,j) \notin S}}{\sum} \frac{A(S \cup \left\{(u,i,j)\}\right) - A(S)}{\binom{|D^{T}|-1}{k}} 
\label{equ:proof11}
\end{aligned}
\end{equation}

It can be easily noted that:
\begin{equation}
\begin{aligned}
    &\frac{1}{|D^{T}|} \sum^{|D^{T}|-1}_{k=0} \underset{|S|=k, (u,i,j) \notin S}{\sum} \binom{|D^{T}|-1}{k}^{-1} \\ &= \frac{1}{|D^{T}|} \sum^{|D^{T}|-1}_{k=0} \binom{|D^{T}|-1}{k} \cdot \binom{|D^{T}|-1}{k}^{-1} \\
    &= \frac{1}{|D^{T}|} \sum^{|D^{T}|-1}_{k=0} = 1
\end{aligned}
\end{equation}

To simplify the notation, let $V^{TS, \pi}_{(u,i,j)} = A(P^{\pi}_{(u,i,j)} \cup \left\{(u,i,j)\}\right) - A(P^{\pi}_{(u,i,j)})$. Then, from the perspective of joining process, the following transformation for the right hand side (RHS) of Eq.~\ref{equ:the1} holds:
\begin{equation}
\begin{aligned}
    &RHS = \frac{1}{|D^{T}|!} \underset{\pi \in \Pi^{T}}{\sum} V^{TS, \pi}_{(u,i,j)} \\
    &= \frac{1}{|D^{T}|!} \sum^{|D^{T}|-1}_{k=0} \underset{|P^{\pi}_{(u,i,j)}|=k}{\sum} \; \underset{G(P^{\pi}_{(u,i,j)})}{\sum} \; \underset{G(F^{\pi}_{(u,i,j)})}{\sum} V^{TS, \pi}_{(u,i,j)}\\
    &= \frac{1}{|D^{T}|!} \sum^{|D^{T}|-1}_{k=0} \underset{|P^{\pi}_{(u,i,j)}|=k}{\sum} k!  
    (|D^{T}|-k-1)! V^{TS, \pi}_{(u,i,j)}\\
    &= \frac{1}{|D^{T}|} \sum^{|D^{T}|-1}_{k=0} \underset{|P^{\pi}_{(u,i,j)}|=k}{\sum} \frac{k!  
    (|D^{T}|- k -1)!}{(|D^{T}|-1)!} V^{TS, \pi}_{(u,i,j)}\\
    &= \frac{1}{|D^{T}|} \sum^{|D^{T}|-1}_{k=0} \underset{|P^{\pi}_{(u,i,j)}|=k}{\sum} \binom{|D^{T}|-1}{k}^{-1} V^{TS, \pi}_{(u,i,j)} , 
\label{equ:proof12}
\end{aligned}
\end{equation}
where $F^{\pi}_{(u,i,j)}$ denotes the set of triplets that follow $(u,i,j)$ in the permutation $\pi \in \Pi^{T}$. $G(\cdot)$ indicates the permutation group which contains all possible permutations consisted of the same triplet set. Comparing Eq.~\ref{equ:proof11} and Eq.~\ref{equ:proof12}, we prove that Eq.~\ref{equ:the1} holds.
\end{proof}

\subsection{Proof for Theorem 2}
We provide the proof for Theorem~\ref{theo2} as follows:
\begin{proof}
Let $U(\pi)$ be the uniform distribution of $\pi$ over all $|D^{T}|!$ permutations of $\Pi(D^{T})$. Each permutation has the probability of $\frac{1}{|D^{T}|!}$ to be sampled. The permutation samples $\pi_1, \pi_2, ..., \pi_n$ for conducting the Monte Carlo approximation are sampled from this distribution. So, the expectation for the estimated triplet Shapley is:
\begin{equation}
\begin{aligned}
    E(\overline{V^{TS, \pi}_{(u,i,j)}}) = \underset{\pi \sim U(\pi)}{E} \frac{1}{n} \underset{k=1}{\sum^n} A(P^{\pi_k}_{(u,i,j)} \cup \left\{(u,i,j)\}\right) - A(P^{\pi_k}_{(u,i,j)})
\label{equ:proof21}
\end{aligned}
\end{equation}
Then, the following equation holds:
\begin{equation} 
\begin{aligned}
    E(\overline{V^{TS, \pi}_{(u,i,j)}}) &=  \frac{1}{n} \underset{k=1}{\sum^n} \underset{\pi_k \sim U(\pi)}{E} A(P^{\pi_k}_{(u,i,j)} \cup \left\{(u,i,j)\}\right) - A(P^{\pi_k}_{(u,i,j)})\\
    &= \underset{\pi \sim U(\pi)}{E} A(P^{\pi}_{(u,i,j)} \cup \left\{(u,i,j)\}\right) - A(P^{\pi}_{(u,i,j)}) \\
    & = \frac{1}{|D^{T}|!} \underset{\pi \in \Pi^{T}}{\sum} A(P^{\pi}_{(u,i,j)} \cup \left\{(u,i,j)\}\right) - A(P^{\pi}_{(u,i,j)}) \\
    & = V^{TS}_{(u,i,j)}
\label{equ:proof22}
\end{aligned}
\end{equation}
Thus, the unbiasedness of Monte Carlo Approximation is proved.
\end{proof}

\subsection{Proof for Theorem 3}
The  proof for Theorem~\ref{theo3} is provided as follows: 
\begin{proof}
    Note that when the triplet $(u,i,j)$ appears at the $k$-th position of the permutation, its preceded triplet set $P^{\pi}_{(u,i,j)}$ has $\binom{|D^{T}|-1}{k-1}$ possibilities. For each $P^{\pi}_{(u,i,j)}$, the $G(P^{\pi}_{(u,i,j)})$ includes $(k-1)!$ different possible permutations consisted of the same triplet set. Meanwhile, the $G(F^{\pi}_{(u,i,j)})$ includes $(|D^{T}|-k)!$ different possible permutations. According to the procedure of $\Omega^{*}$ and Eq.~\ref{equ:the31}, nor matter the specific triplets in $P^{\pi}_{(u,i,j)}$ and the specific order of the permutation that precedes or follows $(u,i,j)$, the $V^{TS, *}_{(u,i,j)}$ keeps the same. To simplify notation, let $ V^{TS, *, \pi}_{(u,i,j)} = A^{*}(P^{\pi}_{(u,i,j)} \cup \left\{(u,i,j)\}\right) - A^{*}(P^{\pi}_{(u,i,j)})$. In the above $|D^{T}|$ permutations, we denote the permutation in which $(u,i,j)$ appears at the $k$-th position by $\pi_k$. Thus, following Eq.~\ref{equ:the31}, we have:
\begin{equation}
\begin{aligned}
    V^{TS, *}_{(u,i,j)} &= \frac{1}{|D^{T}|!} \underset{k=1}{\sum^{|D^{T}|}} \; \underset{|P^{\pi}_{(u,i,j)}|=k-1}{\sum} \; \underset{G(P^{\pi}_{(u,i,j)})}{\sum} \; \underset{G(F^{\pi}_{(u,i,j)})}{\sum} V^{TS, *, \pi_k}_{(u,i,j)} \\
    & = \frac{1}{|D^{T}|!} \underset{k=1}{\sum^{|D^{T}|}} \binom{|D^{T}|-1}{k-1} (k-1)!(|D^{T}|-k)! V^{TS, *, \pi_k}_{(u,i,j)} \\
    & = \frac{1}{|D^{T}|} \underset{k=1}{\sum^{|D^{T}|}} A^{*}(P^{\pi_k}_{(u,i,j)} \cup \left\{(u,i,j)\}\right) - A^{*}(P^{\pi_k}_{(u,i,j)})
\end{aligned}
\end{equation}
\end{proof}

\subsection{Time Complexity Analysis}
\label{sec: time complexity analysis}
Time complexity issue has long been a critical concern that limits the wide application of Shapley value-based methods in practical scenarios. Because they often requires considering all possible marginal contributions whose number grows exponentially with the training set size. To dispel this apprehension and prove that our proposed method exhibits the acceptable time efficiency, we conduct the detailed theoretical analysis to the interpretable triplet importance for personalized ranking, especially the triplet Shapley approximation part.

From the Algorithm~\ref{algorithm:mca}, we can notice that if without Monte Carlo approximation and the truncation operation, the original procedure will take $|D^T|!$ times of the inner loop and each inner loop will conduct the gradient descent for $|D^T|$ times. Besides, the control variates technique will perform the gradient descent for $2$ times in each iteration of inner loop according to $\Omega^*$. Thus, the overall complexity of the original procedure is $\mathcal{O}(|D^T| \cdot |D^T|!)$, which is definitely unacceptable. However, with the introduction of Monte Carlo approximation and the truncation operation, we empirically find that the inner loop will be stopped after executing about $|\mathcal{V}|$ times of iterations and the outer loop also reaches the convergence after iterating for about $|\mathcal{V}|$ times. Therefore, the overall complexity has been reduced to $\mathcal{O}(|\mathcal{V}|^2)$ in practical, greatly lower that the original $\mathcal{O}(|D^T| \cdot |D^T|!)$. 

Besides, it should be mentioned that the Monte Carlo process in Algorithm~\ref{algorithm:mca} (outer loop) is parallerizable up to the number of interactions. Therefore, we can greatly reduce the actual time consumption via deploying our method on multiples machines in parallel. Another way for optimization is to divide items into coarse groups according to their property similarity. Thus, when estimating triplet Shapley, we only conduct the approximation on the (\textit{user}, \textit{positive item group}, \textit{negative item group}) level, which can largely reduce the time consumption while not bring obvious accuracy drop.

\section{Supplementary Experiments}
\subsection{Full Descriptions for Datasets}
The full version of descriptions for the six public datasets in our experiments is as follows:
\begin{itemize}[leftmargin=*]
    \item \textbf{Amazon-Books~\citep{he2016ups}:} This dataset contains book reviews from Amazon. We download the ratings-only dataset and preprocess it like~\citep{wu2022adapting}. In the filtered dataset, each user has reviewed at least 15 books, and each book has been reviewed by at least 20 users. 
    \item \textbf{Amazon-CDs~\citep{he2016ups}:} This is a CD review dataset also from the website above. A similar preprocessing is executed and the filtering thresholds are set as 15 and 20 for users and CDs, respectively.
    \item \textbf{Yelp~\citep{asghar2016yelp}:} This dataset is a subset of Yelp's businesses, reviews, and user data and was originally put together for the Yelp Dataset Challenge. The filtering thresholds are the same as above.
    \item \textbf{Gowalla~\citep{cho2011friendship}:} This dataset was collected worldwide from the Gowalla website (a location-based social networking service). Considering the limited interactions, we only filter the users and items with the thresholds of 2 and 1, respectively.
    \item \textbf{ML-20M~\citep{harper2015movielens}:}   This dataset describes 5-star ratings and free-text tagging activities from MovieLens, an online movie recommendation service. The filtering thresholds are set as 15 and 20 for users and movies, respectively.
    \item \textbf{ML-Latest~\citep{harper2015movielens}:} This is a  recently released dataset from MovieLens. Filtering thresholds are set the same as those in ML-20M.
\end{itemize}
The detailed data statistics of such datasets are provided in Table~\ref{table:dataset}. 

\begin{table}[]
\caption{Data statistics of datasets. The prefixes like Amazon and ML have been omitted due to the page limitation.}
\resizebox{0.48\textwidth}{!}{
\begin{tabular}{|c|c|c|c|c|c|c|}
\hline
Dataset                                                                                                  & Books & CDs & Gowalla & Yelp & 20M & Latest\\ 
User \#                                                                                                   &   33K         &   11K         &   30K      &  52K    &  137K &   216K  \\ \hline
Item \#                                                                                                   &    114K       &    33K         &    41K     & 56K    & 13K &  18K     \\ \hline
Inter \#                                                                                          &       2M     &    467K        &    1M     & 2M   & 20M&   27M    \\ \hline
Density                                                                                                    &      0.063\%      &     0.126\%      &  0.084\%      &  0.069\%   & 1.103\% &  0.681\%    \\ \hline
\end{tabular}}
\label{table:dataset}
\end{table}

\end{document}